\documentclass{article}

\usepackage{latexsym}

\usepackage{fixltx2e}
\usepackage[T1]{fontenc}
\usepackage{ae,aecompl}
\usepackage[utf8]{inputenc}
\usepackage[english]{babel}
\usepackage{verbatim}
\usepackage{graphicx}
\usepackage{mathptmx}
\usepackage{array}
\usepackage{amsfonts}
\usepackage{amsmath}
\usepackage{amsthm}
\usepackage{amssymb}
\usepackage{bm}
\usepackage{mathrsfs}
\usepackage{wrapfig}
\usepackage{stmaryrd}
\usepackage{xcolor}
\usepackage{diagbox}
\usepackage[ruled]{algorithm2e}
\SetInd{0.5em}{0.5em} 

\usepackage{todonotes}
\usepackage{paralist}
\usepackage[nameinlink,capitalise,noabbrev]{cleveref}

\newif\ifcomments

\newcommand{\claudio}[1]{%
\ifcomments %
{\bfseries \scriptsize \color{red} Claudio: #1} %
\else %
\fi%
}%

\newcommand{\casper}[1]{%
\ifcomments %
{\bfseries \scriptsize \color{blue} Casper: #1} %
\else %
\fi%
}%

\newtheorem{example}{Example}
\newcommand{\itr}[2]{\ensuremath{#1^{[#2]}}}
\newcommand{\vect}[1]{\ensuremath{\bm{#1}}}
\newcommand{\comp}[2]{\ensuremath{#1_{#2}}}
\newcommand{\sm}[2]{\ensuremath{#1_{[#2]}}}
\newcommand{\eig}[1]{\ensuremath{\text{Eig}(#1)}}
\newcommand{\real}[1]{\ensuremath{\mathbb{R}\text{e}\{#1\}}}
\newcommand{\Names}{\ensuremath{D}}
\newcommand{\Src}[1]{\ensuremath{S_{[#1]}}}
\newcommand{\order}{\ensuremath{\sigma}}
\newcommand{\uc}{{\vect{u}\vect{c}}}
\newcommand{\aux}{{\vect{aux}}}
\newcommand{\up}{{\vect{u}\vect{p}}}
\newcommand{\Step}{\text{\texttt{doStep}}}
\newcommand{\Rollback}{\text{\texttt{rollback}}}
\newcommand{\Output}{\text{\texttt{getOutput}}}
\newcommand{\unit}[1]{\textit{#1}}
\newcommand{\True}{\text{\texttt{TRUE}}}
\newcommand{\False}{\text{\texttt{FALSE}}}
\newcommand{\converged}{\ensuremath{\mathit{converged}}}
\newcommand{\hasConverged}{\text{\texttt{hasConverged}}}

\newcommand{\brackets}[1]{\ensuremath{ \left[ #1 \right] }}

\newcommand{\set}[1]{\ensuremath{ \left\{ #1 \right\}}}

\newcommand{\pargroup}[1]{\ensuremath{ \left( #1 \right)}}

\newcommand{\dert}[1]{\ensuremath{ \dot{#1} }}
\newcommand{\ddert}[1]{\ensuremath{ \ddot{#1} }}
\newcommand{\partialder}[2]{\ensuremath{ \frac{\partial#1}{\partial#2} }}
\newcommand{\setreal}[0]{\ensuremath{\mathbb{R}}}
\newcommand{\setnat}[0]{\ensuremath{\mathbb{N}}}
\newcommand{\norm}[1]{\left\lVert#1\right\rVert}

\newcommand{\abs}[1]{\left|#1\right|}

\newcommand{\bigO}[1]{\ensuremath{ \mathcal{O}\left( #1 \right)}}
\newcommand{\vectorOne}[1]{\brackets{%
\begin{matrix}
  #1
 \end{matrix}%
}}
\newcommand{\vectorTwo}[2]{\brackets{%
\begin{matrix}
  #1 \\
  #2
 \end{matrix}%
}}

\newcommand{\vectorFour}[4]{\brackets{%
\begin{matrix}
  #1 \\
  #2 \\
  #3 \\
  #4
 \end{matrix}%
}}

\newenvironment{aligneq*}%
{
\begin{equation*}
\begin{aligned}
}{
\end{aligned}
\end{equation*}
}

\newenvironment{aligneq}%
{
\begin{equation}
\begin{aligned}
}{
\end{aligned}
\end{equation}
}

\begin{document}

\title{Co-simulation of Continuous Systems: A Tutorial}

\author{Cl\'{a}udio Gomes \\ [7pt]
University of Antwerp,  \\
claudio.gomes@uantwerpen.be \\
\and
Casper Thule \\[7pt]
Aarhus University \\
casper.thule@eng.au.dk \\
\and
\\ 
Peter Gorm Larsen \\ [7pt]
Aarhus University \\
pgl@eng.au.dk \\
\and
\\
Joachim Denil \\ [7pt]
University of Antwerp, Flanders Make  \\
joachim.denil@uantwerpen.be \\
\and
\\
Hans Vangheluwe \\ [7pt]
University of Antwerp, Flanders Make \\
hans.vangheluwe@uantwerp.be
}

\maketitle



\section{Introduction}

Truly complex engineered systems that integrate physical, software and network aspects are emerging \cite{Nielsen2015}, posing challenges in their design, operation, and maintenance.

The design of such systems, due to market pressure, has to be concurrent and distributed, that is, divided between different teams and/or external suppliers, each in its own domain and each with its own tools \cite{Vangheluwe2002}.
Each participant develops a partial solution, that needs to be integrated with all the other partial solutions. 
The later in the process the integration is done, the higher its cost \cite{Plateaux2009}.
Ideally, the solutions developed independently should be integrated sooner and more frequently, in so-called full system analysis \cite{VanderAuweraer2013}.

Modeling and simulation has improved the development of the partial solutions, but falls short in fostering this holistic development process \cite{Blochwitz2011}.
To understand why, one has to observe that:
\begin{inparaenum}[(i)]
\item models of each partial solution cannot be exchanged or integrated easily, because these are likely developed by a specialized tool;
\item externally supplied models may have Intellectual Property (IP) that cannot be cheaply disclosed to system integrators;
\item as solutions are refined, the system should be evaluated by integrating physical prototypes, software components, and even human operators, in what are denoted as Model/Software/Hardware/Human-in-the-loop simulations \cite{AlvarezCabrera2011}; and
\item the models of each partial solution have different characteristics that can be exploited to more efficiently simulate them, making it difficult to find a technique that fits all kinds of models.
\end{inparaenum}

\emph{Co-simulation} is a generalized form of simulation, where a coupled system is simulated through the composition of simulation units \cite{Gomes2018,Hafner2017,Palensky2017}.
Each unit is broadly defined as a \emph{black box} capable of exhibiting behaviour, consuming inputs and producing outputs, over simulated time.

Many of the problems occurring in co-simulations are due to the ill composition of simulation units that represent continuous systems \cite{Gomes2017}.
As such, we argue that having a basic knowledge of numerical (co-)simulation can help practitioners debug, and even improve the performance of, existing co-simulations.

In this tutorial, we aim to provide the reader with a basic understanding of numerical algorithms, and we show how attempting to simulate an heterogeneous system naturally leads to co-simulation.
Upon completion, the reader should know the many different possible co-simulation approaches, the main concepts involved, and what their tradeoffs are.
Furthermore, the reader will be equipped to understand the more advanced concepts in the co-simulation literature.

The next section provides a top-down overview of all the concepts that will be discussed here.
This concept map will be revisited in all other sections.
In the sections after, each concept will be discussed, in a bottom up manner, so as to increase the complexity gradually.

\section{Main Concepts}

In this section, we will provide an informal top-down overview on the concepts related to co-simulation.
To that end, we will use a \emph{feature model} \cite{Kang1990}: an intuitive diagram that breaks down the main concepts in a domain.
Some of these concepts will only become clear in later sections, as we delve into the details, so we recommend the reader to come back to this section to place these in the grand scheme of things.
More rigorous definitions are given in \cite{Gomes2018}.

First, we summarize the \emph{objective} of running a co-simulation: to reproduce, as accurately as possible, the behavior of a \emph{system under study}.

\begin{figure}[tbh]
\begin{center}
  \includegraphics[width=0.9\textwidth]{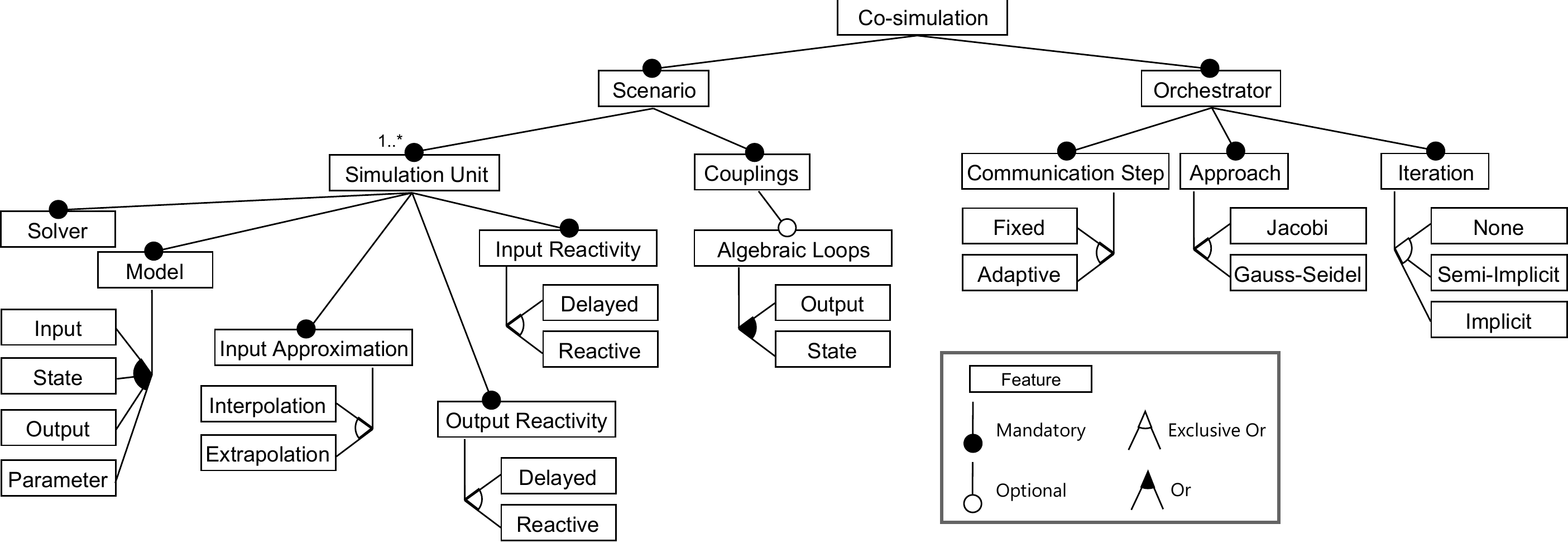}
  \caption{Co-simulation concept breakdown.}
  \label{fig:concept_breakdown}
\end{center}
\end{figure}

\Cref{fig:concept_breakdown} breaks down the main concepts in the co-simulation domain.
To run a co-simulation, one needs a co-simulation scenario and an orchestrator algorithm.

The \emph{co-simulation scenario} points to one or more
simulation units, describes how the inputs and outputs of their models are
related, and includes the configuration of relevant parameters.
 
Each \emph{simulation unit} represents a black box capable of producing behavior.
To produce behavior, the simulation unit needs to have a notion of:
\begin{compactitem}
\item a \emph{model}, created by the modeller based on his knowledge of the system under study;
\item a \emph{solver}, which is part of the modeling tool used by the
modeller, that approximates the behavior of the model; and
\item an \emph{input
approximation}, which approximates the inputs of the model over time, to be
used by the solver; 
\item \emph{input reactivity} and \emph{output reactivity}, that determine which inputs the simulation unit receives from the orchestrator.
\end{compactitem}
 The \emph{orchestrator} is responsible for running the
co-simulation.
 It initializes all the simulation units with the appropriate
values, sets/gets their inputs/outputs, and coordinates their progression over
the simulated time.
 To progress the co-simulation, the orchestrator, after
setting the appropriate inputs to the simulation units (computed from their
outputs according to the co-simulation scenario), asks them to simulate for a
given interval of simulated time, by providing them with a \emph{communication
  step}.
 The simulation units in turn will approximate the behavior of their
model within the interval between the current simulated time and the next
communication time, relying only on the inputs they have received at the previous communication times.
In order to simplify the explanations and analyses presented later in this document, we assume that the simulation units will only receive more inputs at the next communication with the orchestrator,
hence they must rely on their input approximations.

\begin{figure}[tbh]
\begin{center}
  \includegraphics[width=0.8\textwidth]{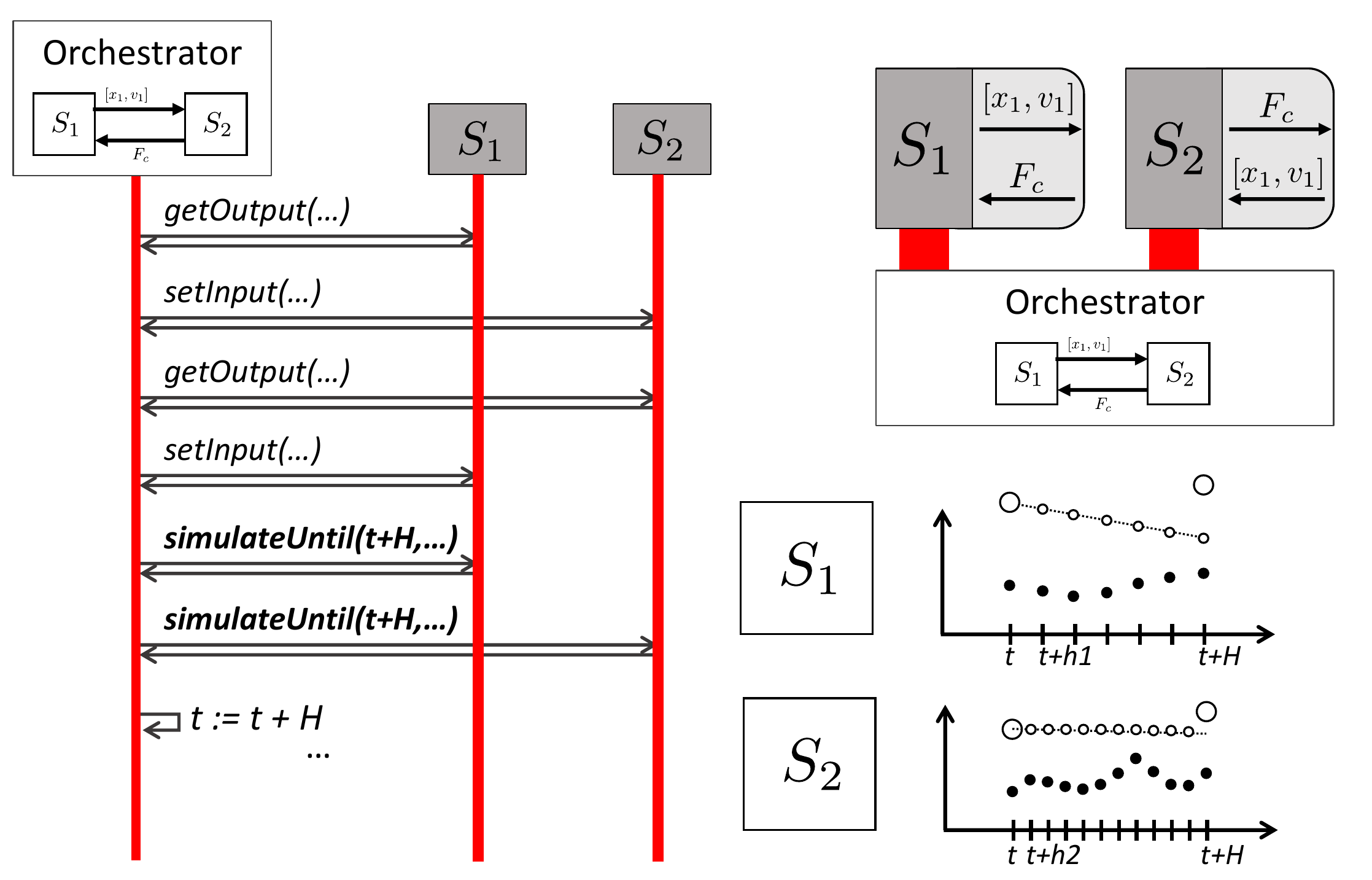}
  \caption{Example co-simulation coordination (left), co-simulation scenario (top right), and internal behavior of simulation units (bottom right).}
  \label{fig:example_coordination}
\end{center}
\end{figure}

\Cref{fig:example_coordination} gives an illustration of these concepts. The
figure in the left-hand side illustrates how the orchestrator coordinates the
co-simulation by getting outputs, setting inputs, and requesting the simulation
units S1 and S2 to progress in time. The figure in the top-right-hand side presents the
co-simulation scenario, where $S_1$ receives input $F_c$ and outputs $[x_1, v_1]$, and $S_2$
receives inputs $[x_1, v_1]$ and outputs $F_c$. The two figures in the
bottom-right-hand side presents the internal behaviour of the simulation units.
The large unfilled dots represent input values, and the smaller unfilled dots
represent their extrapolations, as computed by the simulation units. 
One can see that there is a
difference between the values calculated by the extrapolation functions opposed
to the actual input, due to the gap between the larger and smaller unfilled dots
at $t+H$. 
The black dots represents outputs. As illustrated, $S_1$ and
$S_2$ perform small steps of respectively $h_1$ and $h_2$ internally, until
the time $t+H$ is reached.
 
The communication step size can either be \emph{fixed} (defined before the co-simulation starts and constant throughout its execution), or \emph{adaptive} (the orchestrator determines the best value to be used whenever it asks the simulation units to compute).

\begin{figure}[tbh]
\begin{center}
  \includegraphics[width=0.9\textwidth]{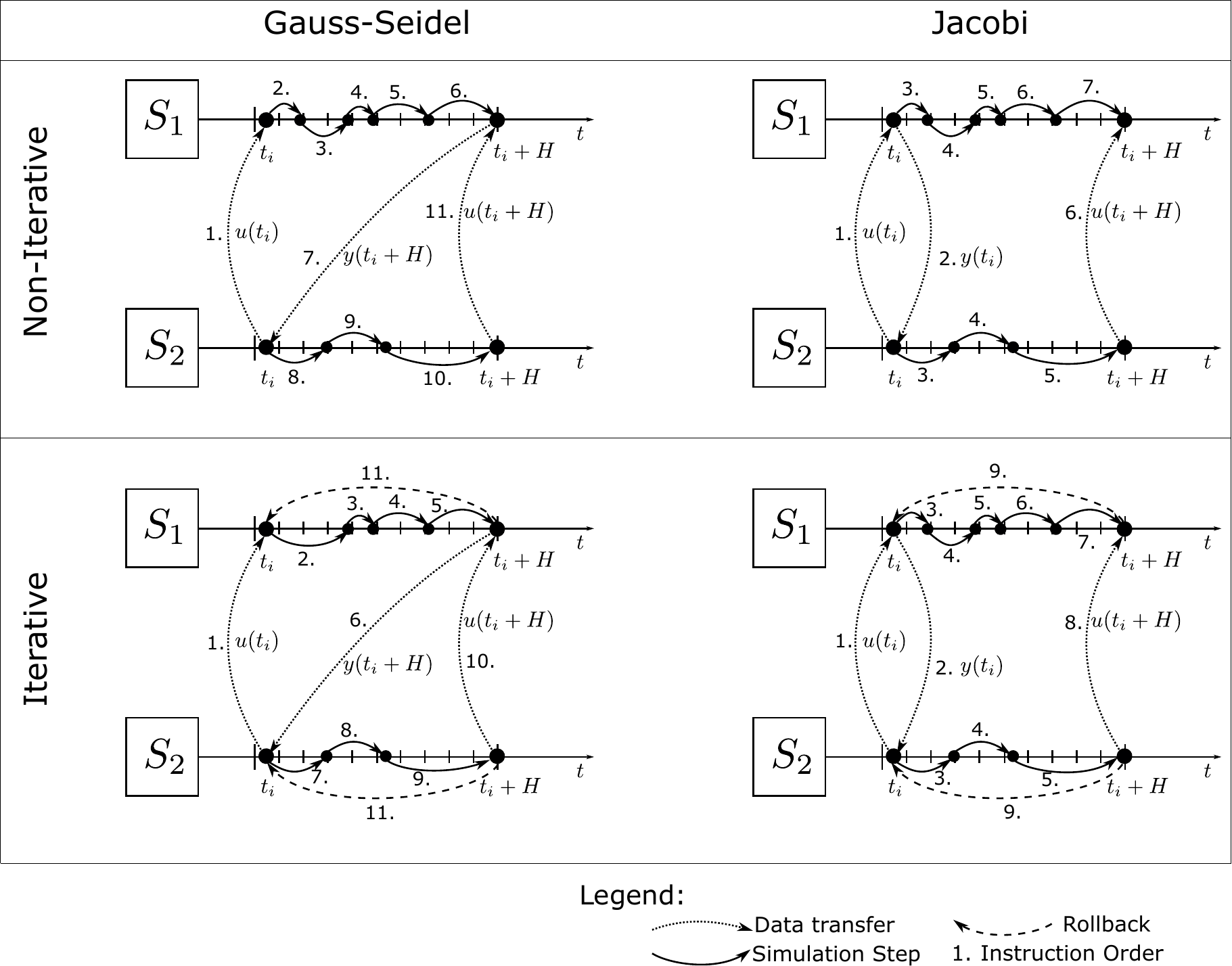}
  \caption{Overview of orchestration algorithms.}
  \label{fig:cosim_overview}
\end{center}
\end{figure}

The \emph{communication approach} encodes the order in which the simulation units are given inputs and instructed to compute the next interval.
\Cref{fig:cosim_overview} summarizes the multiple types of orchestration algorithms using time diagrams.

In the \emph{Gauss-seidel} approach, the orchestrator asks one simulation unit at a time to compute to the next interval and produce outputs. Then, the orchestrator uses the most recent outputs when asking the next unit to compute.

In the \emph{Jacobi} approach, the orchestrator asks all units to compute the interval in parallel, setting their inputs at the end of the co-simulation step.

Finally, the orchestrator may retry the co-simulation step, using improved input estimates, computed from the most recent outputs.
This process can be repeated until there is no improvement on the inputs (\emph{fully implicit iteration}), or a fixed number of iterations has been done (\emph{semi-implicit iteration}).
In the later sections it will become clear why it is a good idea to retry the co-simulation step.

In the following sections, we will follow a bottom up approach, starting with the simplest concepts in \cref{fig:concept_breakdown}, and building our way up to co-simulation.


\section{Models, Solvers, and Input Approximations}
\label{sec:sim_units}

Since co-simulation is a form of generalized simulation, it is paramount that simulation is well understood.
In this section, we cover the basic algorithms to approximate the solution, $x(t)$, of first order Ordinary Differential Equations (ODEs), $\dert{x} = f(x, u)$, having an initial condition, $x(0)=x_0$.
We start with scalar differential equations and then move to vector equations.
A running example will be incrementally constructed, so that the numerical methods introduced can be tried out.

The relationship between the concepts learned in \cref{sec:scalar_ivp,sec:vector_ivps}, and the concept of simulation unit (recall \cref{fig:concept_breakdown}), is discussed in \cref{sec:rel_sim_units}.

\subsection{Scalar Initial Value Problems}
\label{sec:scalar_ivp}

A \emph{scalar Initial Value Problem (IVP)} is defined as a scalar ODE, with an initial condition.
Formally, it has the form:
\begin{equation}\label{eq:scalar_ivp}
\dert{x} = f(x, u) \text{, with } x(0)=x_0,
\end{equation}
where $x: \setreal \to \setreal$ denotes the (scalar) state function, $\dert{x}$ denotes the time derivative of $x$, $f: \setreal^2 \to \setreal$ is a scalar function, $u: \setreal \to \setreal$ is the input function, and $x_0 \in \setreal$ is a given initial value of $x(t)$.

\begin{example}\label{ex:car}
Consider a car whose acceleration is set by a cruise controller, and moves in a straight line.
Let $v(t)$ denote the speed of the car over time, $m$ its mass, and $v_d$ the desired speed (input); and assume that the car is initially moving at speed $v_0$.
Then the scalar IVP is given by
\begin{equation}\label{eq:car_ivp}
\dert{v} = \frac{1}{m} \brackets{k(v_d - v) - c_f v} \text{, with } v(0)=v_0,
\end{equation}
where $k(v_d - v)$ is the acceleration set by the cruise controller, $v_d$ is the input, $k>0$ is the acceleration multiplier constant, and $c_f>0$ is the friction coefficient.
\end{example}

\begin{figure}[tbh]
\begin{center}
  \includegraphics[width=0.6\textwidth]{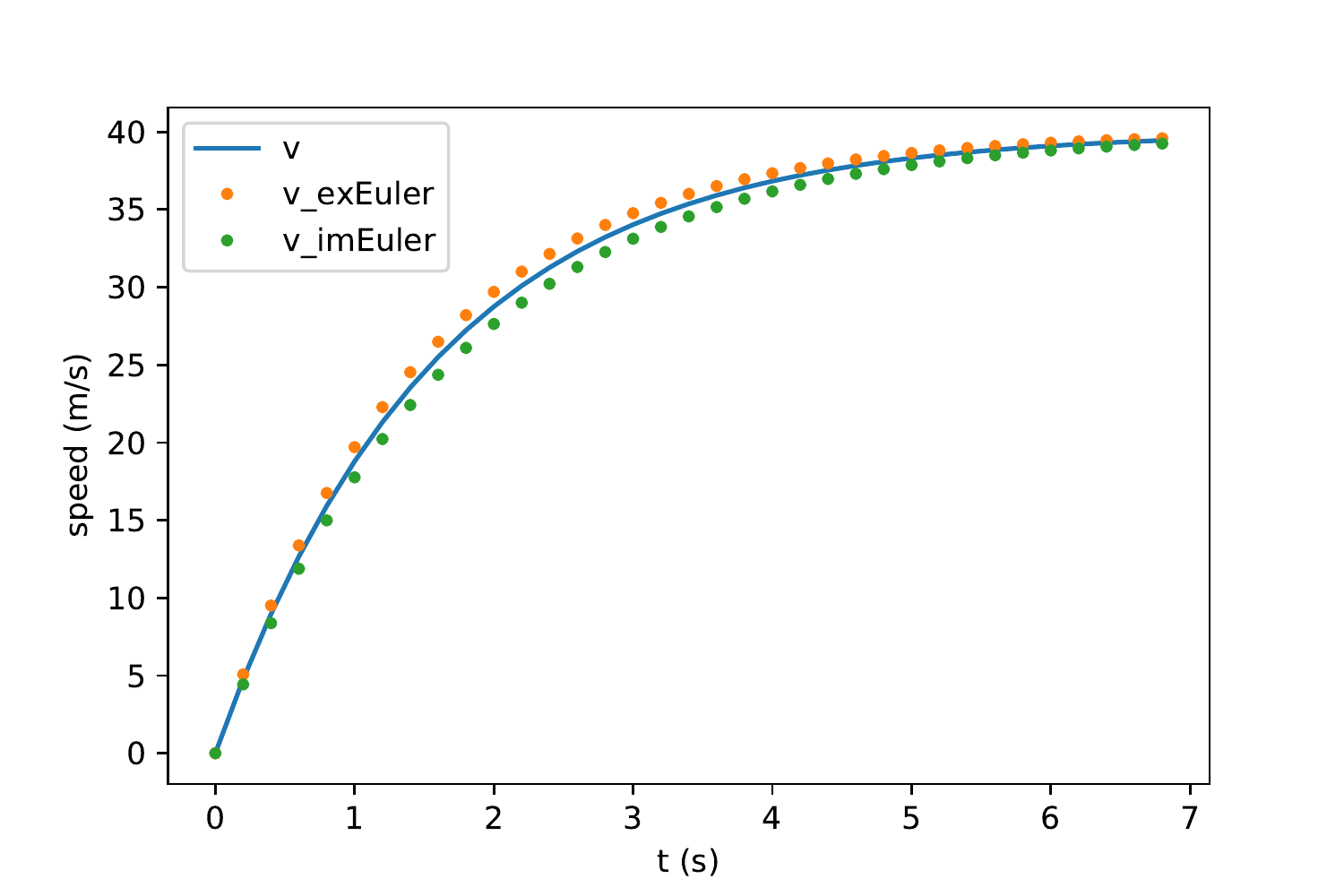}
  \caption{Analytical solution and approximations to the IVP in \cref{ex:car}. Parameters are: $h=0.2, m=1576(kg), v_d=40(m/s), v_0=0(m/s), k=10^3, c_f=0.5$.}
  \label{fig:car_example}
\end{center}
\end{figure}

The solution of the scalar IVP (\ref{eq:scalar_ivp}) is a function $x(t): \setreal \to \setreal$ whose derivative satisfies \cref{eq:scalar_ivp}.
For example, the solution of the IVP posed in the car example (\cref{ex:car}), and plotted in \cref{fig:car_example}, is:
$$
v(t) = \frac{k v_d}{c_f+k} - \pargroup{\frac{k v_d}{m} - v_0}e^{- \frac{t}{m}(c_f+k)}.
$$

In general, it is not possible, nor feasible, to find an explicit solution to the IVP.
Instead, an approximate solution can be computed using a numerical method.
In the following, we provide an intuitive derivation of two variations of Euler's method.

To derive an approximation $\tilde{x}(t)$ of the solution to the scalar IVP in \cref{eq:scalar_ivp}, we start by noting that the initial point is given by the initial value, that is, $\tilde{x}(0) = x_0$, so at least one point is known.
For a small $h>0$, the limit definition of the derivative in the left hand side of \cref{eq:scalar_ivp} can be replaced by its approximation 
$\dert{x} \approx (x(t+h)-x(t))/h$.
By \cref{eq:scalar_ivp}, we have $(x(t+h)-x(t))/h \approx f(x(t), u(t))$, which can be solved for $x(t+h)$ to give the \emph{Explicit Euler Method}:
\begin{equation}\label{eq:euler_method}
x(t+h) \approx x(t) + f(x(t), u(t))h \text{, with } x(0)=x_0.
\end{equation}

Applying \cref{eq:euler_method} to the initial value, gives the point $\tilde{x}(h)$, which approximates $x(h)$.
The procedure can then be repeated using $\tilde{x}(h)$ to compute $\tilde{x}(2h)$, and so on.
This method can be applied to the car example by combining \cref{eq:car_ivp} and \cref{eq:euler_method}:
\begin{equation}\label{eq:euler_car}
v(t+h)=v(t)+\frac{1}{m} \brackets{k(v_d - v(t)) - c_f v(t)} h
\end{equation}
The approximation calculated by \cref{eq:euler_car} with the parameters: $h=0.2, m=1576(kg), v_d=40(m/s), v_0=0(m/s), k=10^3, c_f=0.5$ is shown in \cref{fig:car_example}, i.e., the speed at time $h=0.2$ is calculated by: 
\begin{equation}\label{eq:euler_car_ex}
v(0.2) = 0+\frac{1}{1576}\brackets{10^3(40-0)-0.5*0}*0.2 \approx 5 \text{, as } v(0)=0
\end{equation}

The implicit variation of Euler's method is derived in the same way as the explicit variation.
The difference is that instead of deriving the method from the approximation $(x(t+h)-x(t))/h \approx f(x(t),u(t))$, we evaluate $f$ at the point $x(t+h)$.
That is, we take the approximation $(x(t+h)-x(t))/h \approx f(x(t+h),u(t+h))$, and rearrange it to get the \emph{Implicit Euler Method}:
\begin{equation}\label{eq:imp_euler_method}
x(t+h) \approx x(t) + f(x(t+h),u(t+h))h \text{, with } x(0)=x_0.
\end{equation}

The value of $x(t+h)$ is the unknown in \cref{eq:imp_euler_method}, and $x(t+h)$ depends on itself, that is, it appears on both sides of the equation, creating an \emph{algebraic loop}.
We now present a simple method to estimate $x(t+h)$ in \cref{eq:imp_euler_method}.

The \emph{direct iteration} method\footnote{The direct iteration method is also known as successive substitution, functional iteration, or fixed point iteration.}, computes the solution to an equation $x = g(x)$ by starting from an initial guess denoted as $\itr{x}{0}$ and evaluating the right hand side with it. Then the result is used for the next evaluation of the right hand side, until two successive evaluations are close enough.
In other words, it computes the iteration
\begin{aligneq}\label{eq:direct_iteration}
\itr{x}{1} &= g(\itr{x}{0}); \hspace{2em} \itr{x}{2} = g(\itr{x}{1}); \hspace{2em} \itr{x}{3} = g(\itr{x}{2}); \ldots \\
           & \ \ \ \ \text{ until } \abs{\itr{x}{i+1}-\itr{x}{i}} < \epsilon, \text{ for small } \epsilon>0.
\end{aligneq}%
\noindent
When applying the direct iteration method as part of a simulation step of the implicit Euler method, a good initial guess $\itr{x}{0}$ can be given by the most recently computed value or by an application of the explicit Euler step.
Formally, at simulation time $t$, $\itr{x}{0}=x(t)$, or $\itr{x}{0}=x(t) + f(x(t),u(t))h$.

\begin{example}
  To demonstrate direct iteration, the implicit Euler method presented in \cref{eq:imp_euler_method} can be applied to \cref{eq:car_ivp} to get: 
  \begin{equation}\label{eq:euler_di_car}
  v(t+h)=v(t)+\frac{1}{m} \brackets{k(v_d - v(t+h)) - c_f v(t+h)} h
  \end{equation}
  \Cref{table:di_imp_euler} presents the results of two steps with \cref{eq:euler_di_car} ($v(0.2)$ and $v(0.4)$), using the implicit euler method with the parameters as in \cref{eq:euler_car_ex}. 
  Each step comprises five iterations of the direct iteration method. 
  The value from the last iteration in the first step is used as the initial guess in the second step. The initial guess in the first step is the result of one explicit Euler step as in \cref{eq:euler_car_ex}: $v(0.2) \approx 5$.
  The values in bold represent the result of the implicit euler step.
\end{example}

\begin{table}[tbh]
  \begin{center}
  \caption{Direct iteration applied to the implicit euler applied to the car example.}
  \label{table:di_imp_euler}
  \begin{tabular}{| c | c | c | c | c | c | c |}
    \hline
    \diagbox[width=2.5cm]{Step}{Iteration} & Initial Guess & 1 & 2 & 3 & 4 & \textbf{5} \\ \hline
    $v(0.2)$ &  5 & 4.4413  & 4.5122  & 4.5032  & 4.5044 & \textbf{4.5042} \\ \hline
    $v(0.4)$ & 4.5042 & 9.0085 & 8.4366 & 8.5092 & 8.5000 & \textbf{8.5012} \\
    \hline
  \end{tabular}
\end{center}
\end{table}

\begin{figure}[tbh]
\begin{center}
  \includegraphics[scale=1]{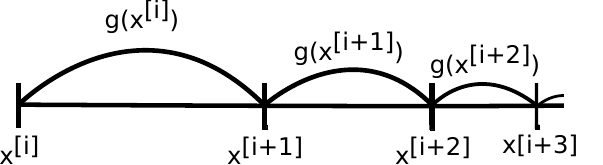}
  \caption{Direct iteration illustration that shows convergence.}
  \label{fig:di}
\end{center}
\end{figure}

The direct iteration method will converge to a solution if successive results get closer and closer, as illustrated in \cref{fig:di}.
Formally, that means that 
\begin{equation}\label{eq:strong_condition}
\abs{g(\itr{x}{i+1})-g(\itr{x}{i})} < \abs{\itr{x}{i+1} - \itr{x}{i}} \Leftrightarrow 
\abs{\frac{g(\itr{x}{i+1})-g(\itr{x}{i})}{\itr{x}{i+1} - \itr{x}{i}}} < 1 \text{ \ \ if } \itr{x}{i+1} - \itr{x}{i} \neq 0
\end{equation}
is satisfied for every $i$.
In the case that $\itr{x}{i+1} - \itr{x}{i} = 0$, then the solution has converged.

\begin{example}
  Applying \cref{eq:strong_condition} to the values in \cref{table:di_imp_euler} results in the values in \cref{table:di_imp_euler_conv}, which shows convergence as every value is smaller than 1. 
\end{example}

\begin{table}[tbh]
  \caption{Convergence applied to \cref{table:di_imp_euler}.}
  \label{table:di_imp_euler_conv}
  \begin{center}
  \begin{tabular}{| c | c | c | c | c |}
    \hline
    \diagbox[width=2.5cm]{Step}{Iteration} & 1 & 2 & 3 & 4 \\ \hline
    $v(0.2)$ &  0.1270 & 0.1270  & 0.1270  & 0.1270  \\ \hline
    $v(0.4)$ & 0.1270 & 0.1270 & 0.1270 & 0.1270  \\
    \hline
  \end{tabular}
\end{center}
\end{table}

Now we derive a condition which is stronger than the above condition, but can be used to predict whether convergence will occur, without actually having to compute the iteration. 
By the Mean Value Theorem, there exists a $\itr{\zeta}{i}$ such that 
$$
\itr{x}{i} < \itr{\zeta}{i} < \itr{x}{i+1} \text{ and } \frac{dg(\itr{\zeta}{i})}{dx} = \frac{g(\itr{x}{i+1})-g(\itr{x}{i})}{\itr{x}{i+1} - \itr{x}{i}}.
$$
The condition
$$
\frac{dg(\itr{\zeta}{i})}{dx} < 1
$$
can be satisfied for all $i$ if we require that 
\begin{equation}\label{eq:convergence_direct_method}
\abs{\frac{dg(x)}{dx}} < 1 \text{, for all } x.
\end{equation}

\cref{eq:convergence_direct_method} shows us that the direct iteration method, when used in combination with the implicit Euler method (\cref{eq:imp_euler_method}), is always guaranteed to converge, provided that the step size $h$ used is small enough.
To see why, let $g(x) = c + f(x,u)h$ denote the direct iteration function, where $c$ and $u$ are known, and $x$ is the unknown.
Differentiating $g$ with respect to $x$, taking the absolute, and adding the restriction in \cref{eq:convergence_direct_method}, yields
\begin{equation}\label{eq:cond_implicit_euler}
\abs{\frac{dg(x)}{dx}} = h \abs{\partialder{f(x,u)}{x}} < 1.
\end{equation}

Applying the above equation to \cref{eq:car_ivp}, with the parameters in \cref{fig:car_example}, yields
$h \abs{\partialder{f(x,u)}{x}} = h \abs{\frac{1}{m}(-c_f -k)} = h*0.63484 < 1$, which means $h$ must satisfy 
$h < 1.5752$.

\Cref{fig:car_example} shows the approximation computed with the Implicit Euler method.

\subsection{Vector Initial Value Problems}
\label{sec:vector_ivps}

In this sub-section, we generalize the numerical techniques introduced in \cref{sec:scalar_ivp} to vector IVPs.
We will denote vectors with bold face, and we will use capital letters for matrices and vector valued functions.
Given a vector $\vect{x}$, we denote its transpose as $\vect{x}^T$.
Furthermore, we denote the $i$-th element of vector $\vect{x}$ by $\comp{x}{i}$, so that $\vect{x} = \vectorOne{\comp{x}{1} & \comp{x}{2} & \cdots & \comp{x}{n}}^T$.
Similarly, $\comp{F}{i}(\vect{x})$ denotes the $i$-th element of the vector returned by $F(\vect{x})$.

An \emph{Initial Value Problem} is the generalization of \cref{eq:scalar_ivp}, to vectors:
\begin{equation}\label{eq:ivp}
\dert{\vect{x}} = F(\vect{x},\vect{u}(t)) \text{, with } \vect{x}(0)=\vect{x_0},
\end{equation}
where $\vect{x}$ and $\vect{u}$ are vector functions, 
and $F$ is a vector valued function.

\begin{figure}[tbh]
\begin{center}
  \includegraphics[width=0.17\textwidth]{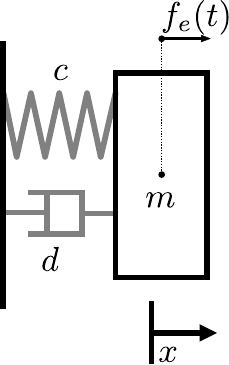}
  \caption{Mass-spring-damper system.}
  \label{fig:msd}
\end{center}
\end{figure}

\begin{example}\label{ex:msd}
The mass-spring-damper system, illustrated in \cref{fig:msd}, is modelled by the following second order ordinary differential equation:
$$
\ddert{x} = \frac{1}{m} (- c x - c_f \dert{x} + f_e(t)),
$$
where $x$ denotes the position of the mass,
$c>0$ is the stiffness coefficient of the spring,
$c_f>0$ is the damping constant of the damper, and 
$f_e(t)$ denotes an external force exerted on the mass.

The above equation can be put into the form of \cref{eq:ivp} by introducing a new variable for velocity, $v = \dert{x}$, and letting the vector $\vect{x} = \vectorOne{x & v}^T$.
Given an initial position $x_0$ and velocity $v_0$,
we obtain the following IVP:
\begin{equation*}
\dert{\vect{x}} = \vectorTwo{\dert{x}}{\dert{v}} = F(\vectorTwo{x}{v}, f_e(t)) = \vectorTwo{v}{(1/m)(- c x - c_f v + f_e(t))} \text{, with } \vect{x}(0) = \vectorTwo{x_0}{v_0}.
\end{equation*}
\end{example}

\begin{figure}[tbh]
\begin{center}
  \includegraphics[width=0.6\textwidth]{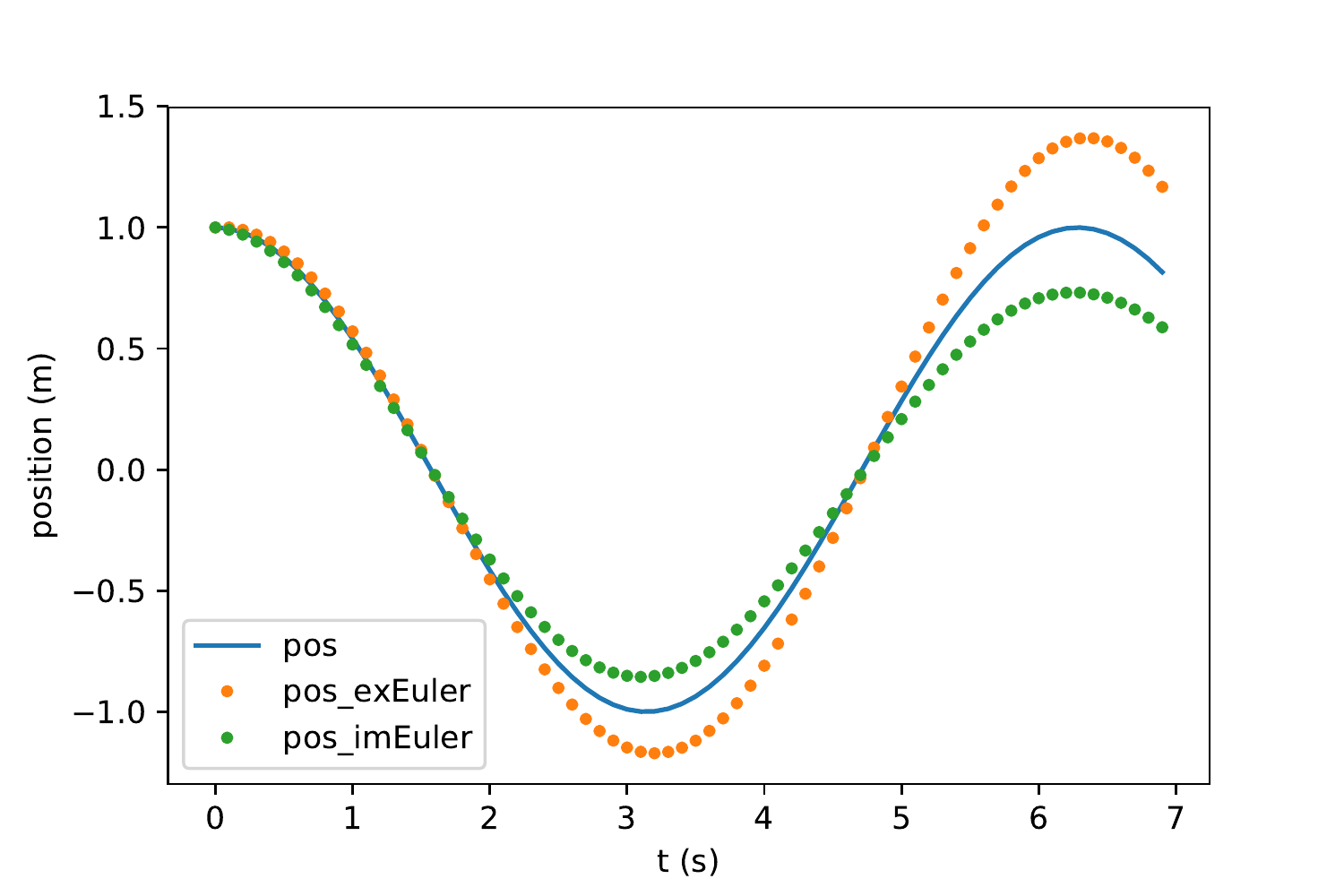}
  \caption{Position (and its approximations) over time of the mass-spring-damper system. Parameters are: $h=0.1,m=c=1,c_f=10^{-4}, f_e(t)=0, \vect{x}_0=\protect\vectorOne{1 & 0}^T$.}
  \label{fig:msd_example}
\end{center}
\end{figure}

The time derivative of a vector is the time derivative of each of its components, so the solution to \cref{eq:ivp} is a vector valued function $\vect{x}(t)$ where each component $\comp{\vect{x}}{i}(t)$ obeys the equation $\comp{\dert{\vect{x}}}{i}(t) = \comp{F}{i}(\vect{x}(t),\vect{u}(t))$, with $\comp{\vect{x}}{i}(0)=\vect{x_{\comp{0}{i}}}$.
As an example, \cref{fig:msd_example} shows the solution of the position component of the mass-spring-damper IVP introduced in \cref{ex:msd}. The solution to the velocity component is omitted.

The \emph{Explicit Euler Method} can be derived as follows.
Taking the limit definition of the derivative of $\comp{x}{i}$, and rearranging as done to obtain \cref{eq:euler_method}, we get
$\comp{x}{i}(t+h) \approx \comp{x}{i}(t) + \comp{F}{i}(\vect{x}(t), \vect{u}(t)) h$, for a small $h>0$.
This equation, applied to all components of $\vect{x}$, can be put in matrix form:
\begin{equation}\label{eq:euler_method_matrix}
\vect{x}(t+h) \approx \vect{x}(t) + F(\vect{x}(t),\vect{u}(t)) h \text{, with } \vect{x}(0)=\vect{x_0}.
\end{equation}

The Implicit Euler Method can be derived analogously:
\begin{equation}\label{eq:implicit_euler_method_matrix}
\vect{x}(t+h) \approx \vect{x}(t) + F(\vect{x}(t+h),\vect{u}(t+h)) h \text{, with } \vect{x}(0)=\vect{x_0}.
\end{equation}

The condition in \cref{eq:cond_implicit_euler} for convergence of the direct iteration (\cref{eq:direct_iteration}) used in combination with the implicit Euler method (\cref{eq:implicit_euler_method_matrix}), also generalizes to the vector IVP (\cref{eq:ivp}) by replacing the absolute $\abs{\cdot}$ by a vector norm $\norm{\cdot}$. It is a special case of the Contraction Mapping Theorem~\cite[Theorem 8.2.2]{Ortega1990}.

\claudio{Help wanted: Provide example of the condition for convergence applied to the implicit euler with the mass spring damper system. Derive the largest step size ensuring convergence.}

\subsection{Constructing Simulation Units}
\label{sec:rel_sim_units}

This subsection describes how the concepts introduced in the previous subsection can be used to construct simulation units.

Models are vector IVPs with output:
\begin{aligneq}\label{eq:model}
\dert{\vect{x}} &= F(\vect{x},\vect{u}) \text{, with } \vect{x}(0)=\vect{x_0}, \text{ and } \\
\vect{y} &= G(\vect{x},\vect{u}),
\end{aligneq}
where $\vect{y}$ denotes the output vector, and $G$ the output function.

Solvers are numerical methods, such as the Euler methods introduced in \cref{eq:euler_method_matrix,eq:implicit_euler_method_matrix}.

To understand the role of input extrapolation functions, we need to recall the interactions between the orchestrator and each simulation unit (recall \cref{fig:example_coordination}).

In order to facilitate the explanation, let us make the following assumptions:
$H>0$ denotes the communication step size, kept the same throughout the co-simulation;
$t_i = i H$ denotes the simulated time at the $i$-th co-simulation step; and
the orchestrator follows a Jacobi approach (see \cref{fig:cosim_overview}).
The other cases should be easy to understand once this one is clear.

Under the above assumptions, the orchestrator, at time $t_i$, constructs the input to the unit, denoted as $u(t_i)$, and then asks the unit to compute until the time $t_{i+1} = t_i + H$.

Between times $t_i$ and $t_{i+1}$, the unit will iteratively approximate the state of the model, only taking into account the inputs $u(t_i),u(t_{i-1}),u(t_{i-2}),\ldots$ that it has received in the past.
As such, the numerical solver employed in the simulation unit is actually solving a modified version of \cref{eq:model}:
\begin{equation}\label{eq:ivp_cosim}
\dert{\vect{x}} = F(\vect{x},\vect{\tilde{u}}(t)) \text{, with } \vect{x}(t_i)=\vect{x_i}, \text{ and } t \in \brackets{t_i, t_{i+1}},
\end{equation}
where $\vect{\tilde{u}}(t)$ is an approximation of $\vect{u}(t)$ in the interval $t \in \brackets{t_i, t_{i+1}}$, built from input samples computed by the orchestrator in the previous co-simulation steps: $\vect{u}(t_i)$, $\vect{u}(t_{i-1})$, $\vect{u}(t_{i-2})$, \ldots. 

In this interval, the goal of the simulation unit is to estimate $\vect{x}(t_{i+1})$, so that the output $\vect{y}(t_{i+1})$ of the model (recall \cref{eq:model}) can be computed and given to the orchestrator.
Since the output $\vect{y}(t_{i+1})$ at time $t_{i+1}$ may depend on the input $\vect{u}(t_{i+1})$ at time $t_{i+1}$, it can be estimated in two ways, depending on the \emph{output reactivity} of the simulation unit:
\begin{compactdesc}
\item[Output reactive:] using the input $\vect{u}(t_{i+1})$ given by the orchestrator, that is,
$$\vect{y}(t_{i+1}) = G(\vect{x}(t_{i+1}),\vect{u}(t_{i+1})),$$
\item[Output delayed:] using the approximation of the input, that is, 
$$\vect{y}(t_{i+1}) = G(\vect{x}(t_{i+1}),\vect{\tilde{u}}(t_{i+1})).$$
\end{compactdesc} 

Regardless of how the output is computed, it can be the case that the numerical method being used internally in the simulation unit from time $t_i$ to $t_{i+1}$, is implemented in a way that requires the availability of the input at time $t_{i+1}$. Formally, this means the state $\vect{x}(t_{i+1})$ at the next communication time is estimated as
\begin{aligneq}\label{eq:simunit_reactive}
\vect{x}(t_{i+1}) &= \delta(\vect{x}(t_i),\vect{u}(t_{i+1}),\vect{u}(t_i),\vect{u}(t_{i-1}),\ldots) \text{, with } \vect{x}(0)=\vect{x_0},
\end{aligneq}%
\noindent%
where $\delta$ encodes the construction of the input extrapolation function, and the iterative application of the numerical method, starting from state $\vect{x}(t_i)$ until state $\vect{x}(t_{i+1})$.
The units employing these methods are denoted as \emph{input reactive}.

In contrast, simulation units are \emph{input delayed} when they do not require the input at time $t_{i+1}$ in order to estimate $\vect{x}(t_{i+1})$:
\begin{aligneq}\label{eq:simunit_in_delayed}
\vect{x}(t_{i+1}) &= \delta(\vect{x}(t_i),\vect{u}(t_i),\vect{u}(t_{i-1}),\ldots) \text{, with } \vect{x}(0)=\vect{x_0}.
\end{aligneq}%

\Cref{tab:types_sim_units} summarizes the types of simulation units.

\begin{table}[tbh]
  \caption{Types of Simulation Units.}
  \label{tab:types_sim_units}
  \begin{center}
  \begin{tabular}{| c | c | c |}
    \hline
    \  & Output Reactive & Output Delayed \\ \hline
    Input Reactive & 
      $
        \begin{matrix}
        \vect{x}(t_{i+1}) = \delta(\vect{x}(t_i),\vect{u}(t_{i+1}),\ldots) \\
        \vect{y}(t_{i+1}) = G(\vect{x}(t_{i+1}),\vect{u}(t_{i+1}))
        \end{matrix}
      $
       &
      $
        \begin{matrix}
        \vect{x}(t_{i+1}) = \delta(\vect{x}(t_i),\vect{u}(t_{i+1}),\ldots) \\
        \vect{y}(t_{i+1}) = G(\vect{x}(t_{i+1}),\vect{u}(t_{i}))
        \end{matrix}
      $
     \\ \hline
    Input Delayed & 
      $
        \begin{matrix}
        \vect{x}(t_{i+1}) = \delta(\vect{x}(t_i),\vect{u}(t_i),\ldots) \\
        \vect{y}(t_{i+1}) = G(\vect{x}(t_{i+1}),\vect{u}(t_{i+1}))
        \end{matrix}
      $
       &
      $
        \begin{matrix}
        \vect{x}(t_{i+1}) = \delta(\vect{x}(t_i),\vect{u}(t_i),\ldots) \\
        \vect{y}(t_{i+1}) = G(\vect{x}(t_{i+1}),\vect{u}(t_{i}))
        \end{matrix}
      $ \\
    \hline
  \end{tabular}
\end{center}
\end{table}

Note that the kind of simulation unit may impose a specific interaction pattern with the orchestrator. 
For example, simulation units that are input reactive cannot be interacted with with a Jacobi approach.
The formal definition of simulation units will be given in \cref{sec:orchestration}, when a more rigorous definition of co-simulation scenario is given.

\subsection{Summary and Further Reading}

This section presented the most basic numerical methods for the simulation of IVPs.
For an introduction to more advanced methods, we recommend \cite{Cellier2006}, and for an in-depth mathematical treatment of these, we recommend \cite{Wanner1991}.
For an overview of modeling with differential equations, see \cite{Cellier1991}.
For an introduction to dynamical systems modeling and simulation, see \cite{vanAmerongen2010}.

Methods for solving equations of the form $x = F(x)$, such as the successive substitution method, are given in, e.g., \cite[Chapter 2]{Burden2010}.
The alternative derivations of the Euler method, see \cite[Section~5.2]{Gomes2016a}.
More details about the derivation of \cref{eq:convergence_direct_method} can be found in \cite[Theorem 2.4]{Burden2010}, and its generalization can be found in \cite[Theorem 8.2.2]{Ortega1990}.
The formal definitions of simulation units, and their types, are based in \cite{Gomes2018a}.

\section{Basics of Co-simulation}

In this section, we show how co-simulation arises naturally out of the need to use specialized numerical methods for different parts of a given IVP.
To exemplify this, we start by introducing a running example.
Then, after showing that it is difficult to simulate the running example using the previously introduced numerical methods, we introduce a new numerical method that can be used to simulate only a part of the example.
Then we introduce co-simulation as a technique that allows the new numerical method to be combined with the previous ones.
Finally, we provide an overview of more advanced techniques that can improve the co-simulation.

\subsection{Motivating Example}

Inspired by the work in \cite{Arcan2003}, we intend to simulate the forces on the body of a passenger in a moving car.
To keep it simple\footnote{For more details about modeling the human body, see \cite[Chapter 5]{Cellier2006} for an introduction, and \cite{Arcan2003}.}, and to combine the models already introduced in \cref{ex:car,ex:msd}, we model the passenger as two coupled mass-spring-damper systems, representing the head and torso, and the vibrations of the motor as a mass-spring system.

\begin{figure}[tbh]
\begin{center}
  \includegraphics[width=0.3\textwidth]{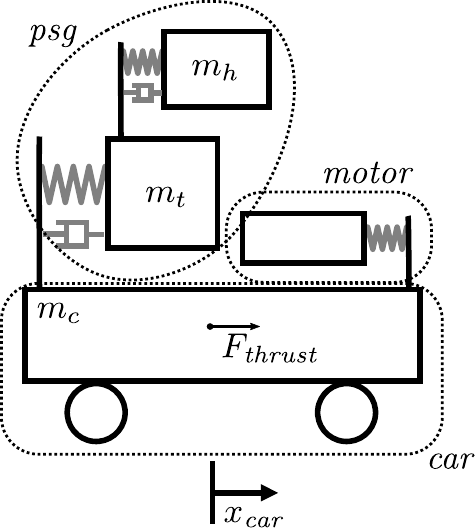}
  \caption{Running example of passenger in an accelerating car with motor vibrations.}
  \label{fig:driver_car}
\end{center}
\end{figure}

\begin{example}\label{ex:car_driver}
The IVP of a passenger in an accelerating car, illustrated in \cref{fig:driver_car}, is given by:
\begin{aligneq}
\mathit{motor}:& \hspace{1em} \ddert{x}_{\mathit{motor}} = -c_m x_\mathit{motor} \\
\mathit{car}:& \hspace{1em} (m_c + m_h + m_b) \ddert{x}_\mathit{car} = k_c (v_d - \dert{x}_\mathit{car}) + a_c x_\mathit{motor} - d_c \dert{x}_{\mathit{car}} \\
\mathit{torso}:& \hspace{1em} m_t \ddert{x}_{\mathit{torso}} = F_h - c_t x_\mathit{torso} - d_t \dert{x}_\mathit{torso} - m_t \ddert{x}_{\mathit{car}} \\
\mathit{head}:& \hspace{1em} m_h \ddert{x}_{\mathit{head}} = - F_h - m_h \ddert{x}_{\mathit{car}} \\
\mathit{coupling\ head\ \&\ torso}:& \hspace{1em} F_h = c_h (x_\mathit{head} - x_\mathit{torso}) + d_h (\dert{x}_\mathit{head} - \dert{x}_\mathit{torso}),
\end{aligneq}
where the initial and parameter values are:
\begin{aligneq}
\mathit{motor}:& \hspace{1em} c_m = 10^4 (m/s), x_\mathit{motor}(0)=1, \dert{x}_\mathit{motor}(0)=0  \\
\mathit{car}:& \hspace{1em} m_c = 1576 (kg), d_c=0.5, k_c=10^3, v_d = 40 (m/s), a_c=5 \times 10^4, \dert{x}_\mathit{car}(0)=0 \\
\mathit{torso}:& \hspace{1em} m_t = 75 (kg), c_t=10^5, d_t=10^5, x_\mathit{torso}(0)=0, \dert{x}_\mathit{torso}(0)=0 \\
\mathit{head}:& \hspace{1em} m_h = 5 (kg), c_h=10^6, d_h=10^4, x_\mathit{head}(0)=0, \dert{x}_\mathit{head}(0)=0
\end{aligneq}
\end{example}

\begin{figure}[tbh]
\begin{center}
  \includegraphics[width=0.5\textwidth]{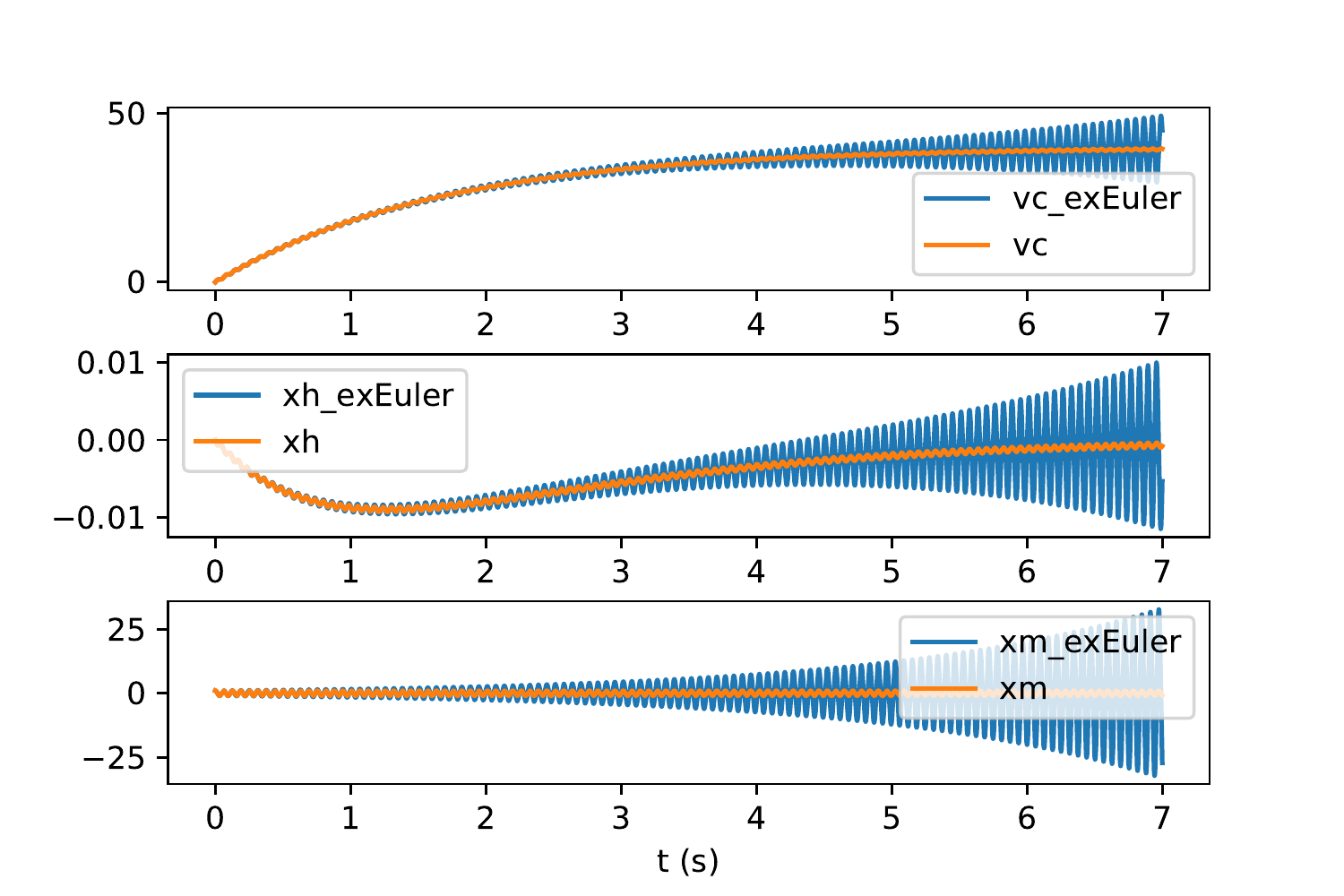}
  \caption{Analytical solution and approximation of the IVP in \cref{ex:car_driver}. The step size $h=10^{-4}s$. The other variables are omitted.}
  \label{fig:car_trace_euler}
\end{center}
\end{figure}

The model introduced in the above example can be put in matrix form as in \cref{eq:linear_forced_ode}. Therefore, its analytical solution can be computed as detailed in \cref{sec:analytical_solution}.
The analytical solution, along its approximation computed with the explicit Euler, is shown in \cref{fig:car_trace_euler}. 

Clearly, the approximation for this example is not satisfactory.
This is because the explicit Euler method is not appropriate to simulate the \unit{motor} sub-system.
To see why, note that the equations governing $x_m$ in \cref{ex:car_driver} are the same as the ones introduced in \cref{ex:msd_simple}, with $c_f = 0$ and $c=c_m$, which means that the Euler method is numerically unstable for this subsystem. 
\casper{This forward reference breaks the flow of reading}
\claudio{I understand. But the problem is that I considered this to be a detour. It's only if the reader wants to understand why\ldots Any idea on how to not make the forward reference?}

\subsection{Specialized Numerical Methods}

The mass-spring is part of a more general class of problems called second derivative IVPs.

\emph{Second derivative IVPs} have the form:
\begin{equation}\label{eq:2ndder_ivp}
\ddert{\vect{x}} = F(\vect{x}, \vect{u}) \text{, with } \vect{x}(0)=\vect{x_0} \text{, and } \dert{\vect{x}}(0)=\vect{v_0},
\end{equation}
and these typically show up in IVPs over equations modeling physical systems where energy is conserved \cite[Chapter 5]{Cellier2006}.

Instead of converting the above IVP to a first order one (as illustrated in \cref{ex:msd}), there are numerical methods that take advantage of the special structure of this problem.

\emph{Godunov’s method} computes the approximated solution to the second derivative IVP in \cref{eq:2ndder_ivp} using the following iteration:
\begin{equation}\label{eq:godunov_method}
\vect{x}(t+h) \approx 2\vect{x}(t) - \vect{x}(t-h) + F(\vect{x}(t),\vect{u}(t))h^2 \text{, with } \vect{x}(0)=\vect{x_0} \text{ and } \vect{x}(h)=\vect{x_h},
\end{equation}
where $\vect{x_h}$ is given.
At time $t$, this method requires access to two previously computed approximations ($\vect{x}(t)$ and $\vect{x}(t-h)$), which it can only be used from $t=2$ onward. 
Fortunately, at this time, the value $\vect{x}(h)$ can be computed using other numerical methods.
Godunov’s method is an example of a \emph{multi-step numerical method} \cite[Chapter III]{Wanner1991}.

As an example, the bottom plot of \cref{fig:cosim_simple} shows the motor vibrations computed with Godunov’s method.

Despite the good performance of Godunov’s method to simulate the mass-spring subsystem of \cref{ex:car_driver}, it cannot be used as is to simulate the complete IVP\footnote{It is possible to adapt Godunov’s method to simulate IVPs that involve the first derivative (see, e.g., \cite[Section 5.5]{Cellier2006}).}, as it does not have the form of \cref{eq:2ndder_ivp}.
In the following sub-sections, we show how to decouple the example into sub-problems, and solve each with the most appropriate numerical method. 

\subsection{Decoupling IVPs}

In the light of the concepts introduced earlier, we can refine the definition of the concepts in \cref{fig:concept_breakdown}.

Here, we define \emph{Co-simulation} as a technique to couple numerical methods, each responsible for a part of the given IVP, in order to approximate the solution to that IVP.

The \emph{configuration} of the co-simulation scenario is an assignment of values to the parameters that affect the co-simulation execution.
For example, one such parameter is the co-simulation step size $H>0$, which controls the points in time at which the numerical methods will exchange values (i.e., at multiples of $H$).
The concrete set of parameters depend on the co-simulation orchestrator so we do not detail them here.
Each \emph{model} represents an IVP, and all models represent a decomposition of the system under study, which we consider to be an IVP as well\footnote{There are examples of co-simulations where the original model is not an IVP, but instead is a differential algebraic system. See \cite{Hafner2013,Busch2011,Kubler2000,Arnold2014}.}.

\begin{figure}[tbh]
\begin{center}
  \includegraphics[width=0.4\textwidth]{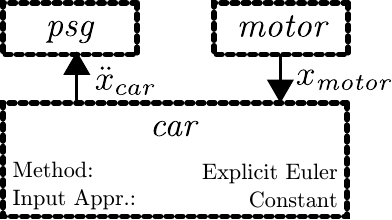}
  \caption{Co-simulation scenario described in \cref{ex:cosim_scenario}.}
  \label{fig:driver_car_cosim}
\end{center}
\end{figure}

\begin{example}\label{ex:cosim_scenario}
An example co-simulation scenario for the IVP introduced in \cref{ex:car_driver} is summarized in \cref{fig:driver_car_cosim}.
The vibrations of the car are approximated using the Godunov’s method (\cref{eq:godunov_method}), and the other two parts are simulated with the Explicit euler method.
\end{example}

\subsection{Orchestration}
\label{sec:orchestration}

The algorithm that processes the co-simulation scenario and coordinates the execution of the simulation units is called the \emph{orchestrator}\footnote{The orchestrator is also known as the master, or coordinator.}.
In this sub-section, we introduce the \emph{Gauss Seidel} and \emph{Jacobi} orchestration algorithms, named after the analogous techniques to solve linear systems.
To explain these methods, we need to first detail the elements that comprise a co-simulation scenario.

Let $H>0$ denote the given communication time step.
We denote the $i$-th communication time as $t_i = i H$.
We say that the $i$-th step of the co-simulation is finished when all the numerical methods have computed their solutions up to, and including, time $t_i$.

Each model is associated with a reference $w \in \Names$, where $\Names$ is a set of all model names.
The model $w$ is an IVP with output:
\begin{aligneq}\label{eq:partition_ivp}
\sm{\dert{\vect{x}}}{w} &= \sm{F}{w}(\sm{\vect{x}}{w},\sm{\vect{u}}{w}) \text{, with } \sm{\vect{x}}{w}(0)=\vect{x_{\sm{0}{w}}}, \text{ and } \\
\sm{\vect{y}}{w} &= \sm{G}{w}(\sm{\vect{x}}{w},\sm{\vect{u}}{w}),
\end{aligneq}
where $\sm{\vect{y}}{w}$ denotes the output vector, and $\sm{G}{w}$ the output function.

As described in \cref{sec:rel_sim_units}, the input function $\sm{\vect{u}}{w}(t)$ is an approximation (i.e., extrapolation or interpolation) constructed from samples of the outputs of other models.
We will denote the set of models whose output is used to construct the input $\sm{\vect{u}}{w}(t)$, as $\Src{w} \subseteq \Names$, standing for \emph{Source models}.
With this notation, for $t \in \brackets{t_i, t_{i+1}}$, the input $\sm{\vect{u}}{w}(t)$ is constructed from the samples of the outputs of every model $v \in \Src{w}$ at the current and previous co-simulation steps. 
The number of samples needed depend on the concrete approximation technique.

We will use $w$ to refer both to the model and the simulation unit, when there is no ambiguity.

Roughly, the task of the orchestrator at time $t_i$ is to provide the output samples that each unit $w$ needs, and ask the unit to approximate the value of $\sm{\vect{y}}{w}(t_{i+1})$.
Therefore, the orchestrator needs to distinguish units according to not only which samples are required to construct their input functions (input reactive or delayed), but also whether their output functions require actual values for inputs or not (output delayed or reactive).
We now describe formally each type of simulation units introduced in \cref{tab:types_sim_units}.

A unit $w$ is \emph{input reactive} if, at any  $t \in \brackets{t_i, t_{i+1}}$, there is at least one $v \in \Src{w}$ such that the input $\sm{\vect{u}}{w}(t)$ depends on the value of $\sm{\vect{y}}{v}$ at time $t_{i+H}$.
Otherwise, $w$ is \emph{input delayed}.

A unit $w$ is \emph{output reactive} if at any time $t_i$, there is at least one $v \in \Src{w}$ such that the computation of the output $\sm{\vect{y}}{w}(t_i)$ requires the value of $\sm{\vect{y}}{v}$ at $t_{i}$.
Otherwise, $w$ is \emph{output delayed}.
Note that the output function $\sm{G}{w}$ may still depend on the input for a unit $w$ that is \emph{output delayed}: it just means that the unit will employ the input approximation in place of the actual input.

The reactivity properties can be seen as contracts between the simulation units and the orchestration algorithm. 
These are specific to how the simulation units are implemented, and not to the sub-models themselves.
In order words, the same sub-model may be implemented differently in different simulation units.

\subsubsection{Gauss-Seidel Orchestrator}
\label{sec:gauss_seidel}

With the above classification, the Gauss-Seidel orchestrator can determine which outputs are used to compute which inputs, and at which times.
This allows it to sort the execution of the units, so that the output samples on which they depend are always available.

At the $i$-th co-simulation step, a unit $w$ \emph{must be executed after} unit $v$ if $v \in \Src{w}$ and $w$ is (input or output) reactive.

To keep the orchestrator simple we assume that the units can always be sorted. In \cref{sec:implicit_orchestration} we relax this assumption. 

We denote the order with a map $\order: \setnat \to \Names$, that returns the unit reference $\order(j)$ that is the $j$-th in the order.
For example, the unit $\order(1)$ is the first.

Under these assumptions and notation, the Gauss-Seidel orchestrator is summarized in \cref{alg:ct_cosim_gauss_seidel}.
Function $C_{w}\pargroup{\set{\sm{\vect{y}}{v} | v \in \Src{w}}}$ computes the input sample of unit $w$ from the output samples of its sources.
The function $\Output(w, \sm{\uc}{w})$ asks unit $w$ to compute the output, optionally using the value in the variable $\sm{\uc}{w}$
Likewise, function $\Step(w, H, \sm{\uc}{w}, \sm{\up}{w})$
asks unit $w$, assumed to be in state $\sm{\vect{x}}{w}(t)$, to compute the value $\sm{\vect{x}}{w}(t+H)$, using either one of the variables provided, depending on its type (i.e., use $\sm{\uc}{w}$ if the unit is input reactive, or use $\sm{\up}{w}$ otherwise).
Any other previous inputs the unit may require are assumed to be stored in its internal state (collected from previous calls to the $\Step$ function.

\begin{algorithm}[htb]
\smaller
\caption{Gauss-seidel orchestrator. See \cref{fig:cosim_overview}.}
\label{alg:ct_cosim_gauss_seidel}
\KwData{The stop time $T$, a communication step size $H$, a co-simulation scenario with unit references $\Names$, and their order $\order$.}
$t := 0$ \tcp*{Simulation time}
\tcp{Initialize variables}
\For{$w \in \Names$}{
  $\sm{\uc}{w} := \sm{\vect{y}}{w} := \vect{0}$ \tcp*{Current I/O variables.}
  $\sm{\up}{w} := \vect{0}$ \tcp*{Previous input variables.}
}
\tcp{Compute initial outputs}
\For{$j = 1,\ldots,\abs{\Names}$}{
  $w := \order(j) $\;
  $\sm{\uc}{w} := C_{w}\pargroup{\set{\sm{\vect{y}}{v} | v \in \Src{w}}} $\tcp*{Compute input from set of sources.}
  $\sm{\vect{y}}{w} :=  \Output(w, \sm{\uc}{w})$\tcp*{Compute output.}
  $\sm{\up}{w} := \sm{\uc}{w}$\;
}
\While{$t < T$}{
  \For{$j = 1,\ldots,\abs{\Names}$}{
    $w := \order(j)$\;
    $\sm{\uc}{w} := C_{w}\pargroup{\set{\sm{\vect{y}}{v} | v \in \Src{w}}} $\;
    $\Step(w, H, \sm{\uc}{w}, \sm{\up}{w})$\tcp*{Compute $\sm{\vect{x}}{w}(t+H)$ from $\sm{\vect{x}}{w}(t)$ and inputs.}
    $\sm{\vect{y}}{w} := \Output(w, \sm{\uc}{w})$\;
  }
  \For{$w \in \Names$}{
    $\sm{\up}{w} := \sm{\uc}{w}$\tcp*{Update previous input.}
  }
  $t := t + H$\tcp*{Advance time}
}
\end{algorithm}

\begin{figure}[tbh]
\begin{center}
  \includegraphics[width=0.8\textwidth]{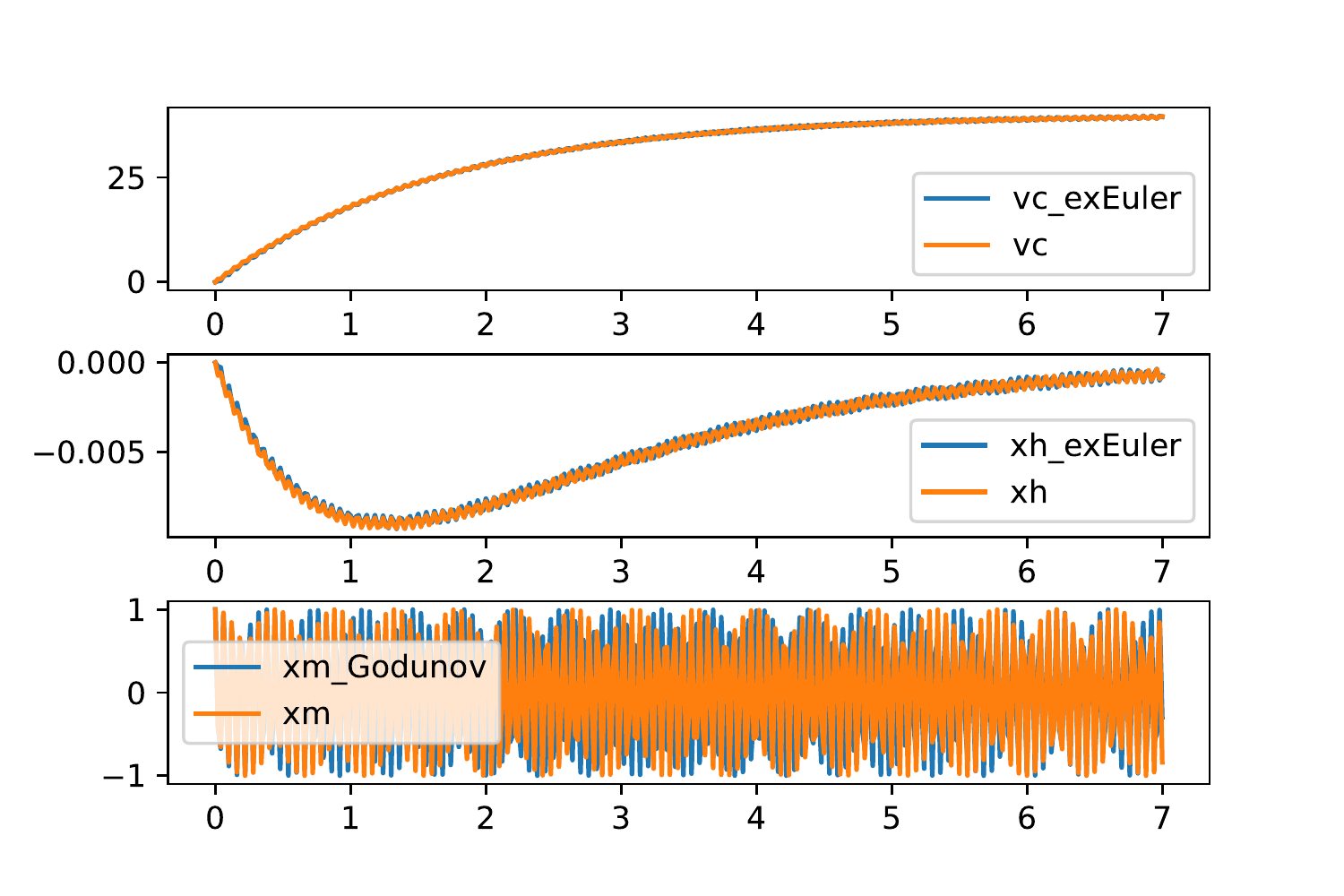}
  \caption{Co-simulation of \cref{ex:cosim_scenario}.}
  \label{fig:cosim_simple}
\end{center}
\end{figure}

\Cref{fig:cosim_simple} shows the solution approximated with the co-simulation of \cref{ex:cosim_scenario} using \cref{alg:ct_cosim_gauss_seidel}.

\subsubsection{Jacobi Orchestrator}

The main difference between the Jacobi and Gauss-Seidel orchestrator lies in the fact that the Jacobi orchestrator assumes that every simulation unit is input delayed.
This has a couple of consequences:
\begin{compactitem}
\item  There is no need to order the units for the execution of the $\Step$ function.
However, the units can still be output reactive/delayed, so the invocations of the $\Output$ functions still need to be sorted.
\item There is no need to keep track of the previous inputs to each unit.
\end{compactitem}

The Jacobi orchestrator is summarized in \cref{alg:ct_cosim_jacobi}.
Compared to the Gauss-seidel orchestrator, the Jacobi is in general less accurate (due to the fact that units cannot use interpolation techniques), but can take advantage of parallelism. 

\begin{algorithm}[htb]
\smaller
\caption{Jacobi orchestrator. See \cref{fig:cosim_overview}.}
\label{alg:ct_cosim_jacobi}
\KwData{The stop time $T$, a communication step size $H$, a co-simulation scenario with unit references $\Names$, and the order $\order$ of their inputs.}
$t := 0$ \tcp*{Simulation time}
\tcp{Initialize variables}
\For{$w \in \Names$}{
  $\sm{\uc}{w} := \sm{\vect{y}}{w} := \vect{0}$ \tcp*{Current I/O variables.}
}
\While{$t < T$}{
  \tcp{Compute outputs in order}
  \For{$j = 1,\ldots,\abs{\Names}$}{
    $w := \order(j)$\;
    $\sm{\uc}{w} := C_{w}\pargroup{\set{\sm{\vect{y}}{v} | v \in \Src{w}}} $\;
    $\sm{\vect{y}}{w} := \Output(w, \sm{\uc}{w})$\;
  }
  \For{$w \in \Names$}{
    $\Step(w, H, \sm{\uc}{w})$\tcp*{Compute $\sm{\vect{x}}{w}(t+H)$ from $\sm{\vect{x}}{w}(t)$ and inputs.}
  }
  $t := t + H$\tcp*{Advance time}
}
\end{algorithm}

\subsubsection{Implicit and Semi-Implicit Orchestrators}
\label{sec:implicit_orchestration}

The Jacobi and Gauss-Seidel orchestration algorithms have iterative counterparts (recall \cref{fig:concept_breakdown}).
An iterative orchestration algorithm will retry each co-simulation step multiple times.
If the number of repetitions is fixed, then we say that the orchestration is \emph{semi-implicit}.
If, on the other hand, the co-simulation step is repeated until some criteria is met, then the orchestration is \emph{implicit}.

In general, iterative techniques are useful when the non-iterative techniques fail to preserve the stability of the original IVP, or when there are algebraic loops in the co-simulation scenario.
When there are algebraic loops, then the units cannot be sorted, as assumed in \cref{sec:gauss_seidel}.

We distinguish two kinds of algebraic loops in co-simulation \cite{Gomes2018,Kubler2000}:
\begin{inparaitem}[\textbullet]
\item \textbf{output loops:} the ones spanning just output variables; and 
\item \textbf{state loops:} the ones that include state variables as well.
\end{inparaitem}
Output loops arise when there is an output of a simulation unit that depends (through the couplings of the co-simulation scenario) on itself, while state loops happen when the state of an input reactive simulation unit depends on itself.

\begin{figure}[tbh]
\begin{center}
  \includegraphics[width=0.8\textwidth]{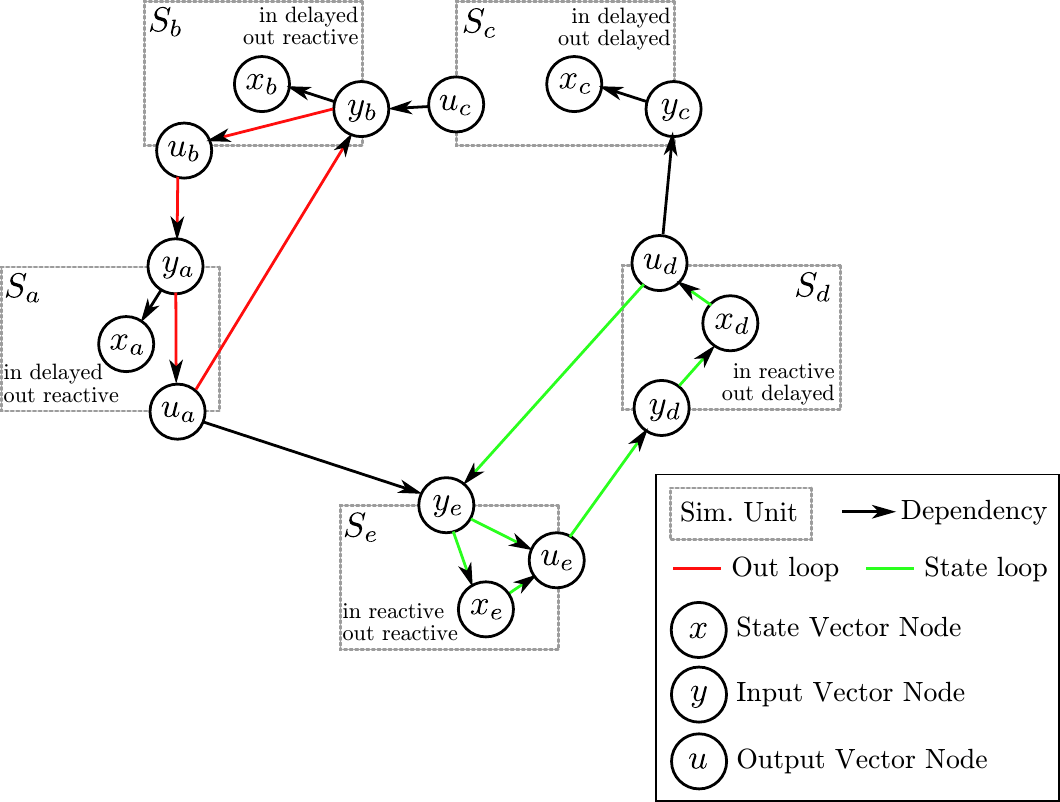}
  \caption{Abstract example co-simulation scenario with the dependency graph. The inputs, state and output variables are vectors. There are multiple algebraic loops.}
  \label{fig:example_dep_graph}
\end{center}
\end{figure}

To illustrate these, we introduce a directed graph based notation to represent dependencies.
Each vector of outputs/inputs/state is represented as one node.
The edges are drawn as follows:
\begin{compactitem}
\item whenever an output vector depends on an input or state vector, an edge is drawn between the corresponding nodes;
\item when an input depends---through the couplings of a co-simulation scenario---on an output, an edge is drawn between the corresponding nodes.
\item when the state evolution function uses an input interpolation approximation (that is, when the unit is input reactive), then an edge is drawn between the input node and the state node.
\end{compactitem}
\Cref{fig:example_dep_graph} shows an abstract example co-simulation scenario illustrating the different dependencies. 

With the dependency relationship introduced, the algebraic loops give rise to cycles in the graph.
If a cycle has nodes that correspond to a state vector, then it is a state loop.
Otherwise, it is an output loop.
These algebraic loops are highlighted in \cref{fig:example_dep_graph}.

\Cref{alg:ct_cosim_gauss_seidel_iterative} illustrates the iterative version of the Gauss-seidel orchestrator.
Function $\hasConverged$ encodes the test for convergence, which can either count a fixed number of iterations (semi-implicit method), or check whether the output values have converged (implicit method).
The $\Rollback$ function reverts the state of the simulation unit to the one before the most recent call to the $\Step$ function.
Contrarily to its non-iterative counterpart, the order used in this algorithm does not necessarily reflect the dependencies between simulation units: it is merely an order defined by the user.

\begin{algorithm}[htb]
\scriptsize
\caption{Iterative Gauss-seidel orchestrator. See \cref{fig:cosim_overview}.}
\label{alg:ct_cosim_gauss_seidel_iterative}
\KwData{The stop time $T$, a communication step $H$, a scenario with unit references $\Names$, and their order $\order$.}
$t := 0$ \tcp*{Simulation time}
\tcp{Initialize variables}
\For{$w \in \Names$}{
  $\sm{\uc}{w} := \sm{\vect{y}}{w} := \vect{0}$ \tcp*{Current I/O variables.}
  $\sm{\up}{w} := \sm{\aux}{w} := \vect{0}$ \tcp*{Previous and auxiliary I/O variables.}
}
\tcp{Compute initial outputs}
$\converged := \False$\;
\While{$t < T$}{
  \For{$j = 1,\ldots,\abs{\Names}$}{
    $w := \order(j) $\;
    $\sm{\uc}{w} := C_{w}\pargroup{\set{\sm{\vect{y}}{v} | v \in \Src{w}}} $\tcp*{Compute input from set of sources.}
    $\sm{\vect{y}}{w} := \Output(w, \sm{\uc}{w})$\tcp*{Compute output.}
    $\sm{\up}{w} := \sm{\uc}{w}$\;
  }
  \uIf{$\hasConverged\pargroup{\set{(\sm{\uc}{w},\sm{\aux}{w}) | w \in \Names}}$}{
    $\converged := \True$\;
  }
  \Else{
    $\sm{\aux}{w} := \sm{\uc}{w}$ for each $w \in \Names$\;
  }
}
\While{$t < T$}{
  $\converged := \False$\;
  \For{$j = 1,\ldots,\abs{\Names}$}{
    $w := \order(j)$\;
    $\sm{\uc}{w} := C_{w}\pargroup{\set{\sm{\vect{y}}{v} | v \in \Src{w}}} $\;
    $\Step(w, H, \sm{\uc}{w}, \sm{\up}{w})$\tcp*{Compute $\sm{\vect{x}}{w}(t+H)$ from $\sm{\vect{x}}{w}(t)$ and inputs.}
    $\sm{\vect{y}}{w} := \Output(w, \sm{\uc}{w})$\;
  }
  \uIf{$\hasConverged\pargroup{\set{(\sm{\uc}{w},\sm{\aux}{w}) | w \in \Names}}$}{
    $\converged := \True$\;
    $\sm{\up}{w} := \sm{\uc}{w}$ for each $w \in \Names$\tcp*{Update previous input.}
  }
  \Else{
    $\sm{\aux}{w} := \sm{\uc}{w}$ for each $w \in \Names$\;
    $\Rollback(w)$ for each $w \in \Names$\tcp*{Cancel the effects of \Step.}
  }
  $t := t + H$\tcp*{Advance time}
}
\end{algorithm}

The iterative version of the Jacobi algorithm is similar, so we omit it.

\subsection{Advanced Co-simulation Techniques}

In order to simplify the explanation, and to keep the algorithms within one page, we made some assumptions in the previous sub-sections.
Additionally, there are more advanced techniques that can be applied in practice to improve the performance of co-simulations.
We discuss these in the following.

\subsubsection{Initialization}

Until now we have assumed that, in a co-simulation scenario, each simulation unit has a given initial state, independent of the initial state of other simulation units.
In practice, this might not be the case, so the co-simulation scenario has to include a description of how the initial states are related (see the \emph{initial state couplings} concept in \cref{fig:concept_breakdown}), and the orchestrator has to compute these initial states.
This computation is similar to the computation of initial outputs, and may also include algebraic loops.

\subsubsection{Fine Grained Input/Output Dependencies}

\begin{figure}[tbh]
\begin{center}
  \includegraphics[width=0.8\textwidth]{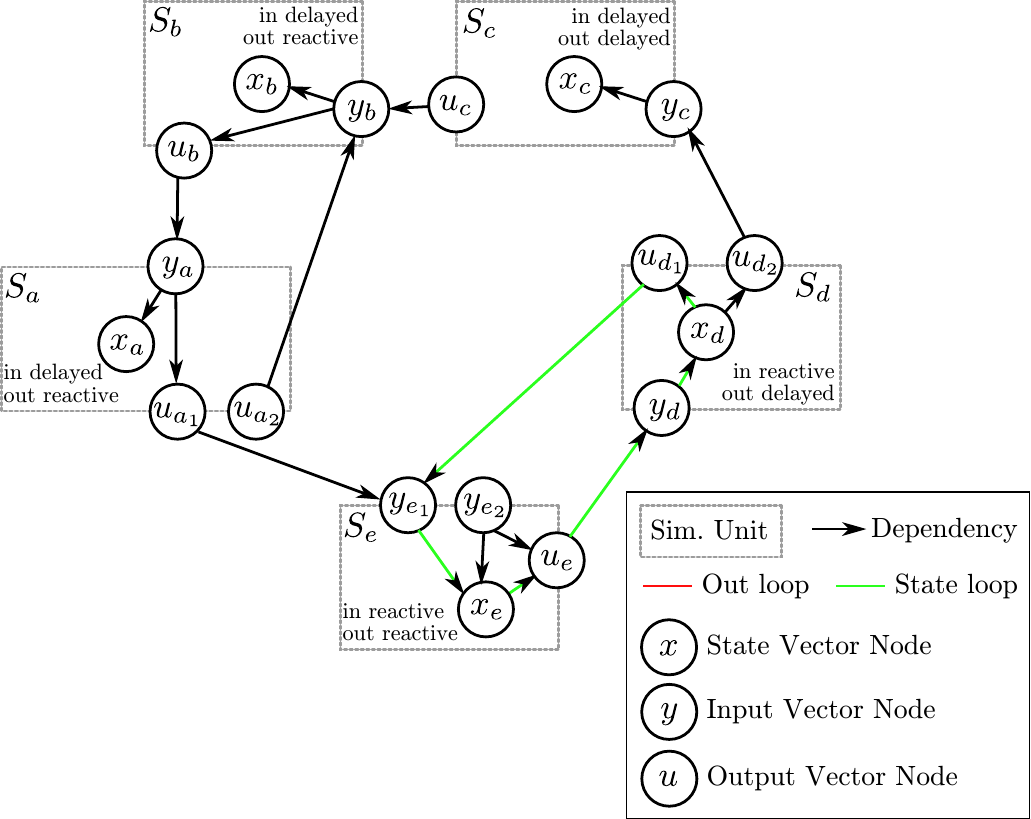}
  \caption{Scalar version of the dependency graph for the co-simulation scenario in \cref{fig:example_dep_graph}. The nodes represent scalar quantities.}
  \label{fig:example_dep_graph_scalar}
\end{center}
\end{figure}

We also assumed that the representation of the dependency information between input, state, and output vectors was adequate.
It is better to represent the dependencies between the scalar variables.
To see why, observe the example in \cref{fig:example_dep_graph_scalar}, which shows the same co-simulation scenario as the one in \cref{fig:example_dep_graph}, but instead of using the vector level dependency information, it uses the the scalar level.
What was identified as an output algebraic loop in \cref{fig:example_dep_graph}, is no longer one at the scalar level dependency graph in \cref{fig:example_dep_graph_scalar}.
This is called a \emph{virtual algebraic loop} and does not require iterative techniques to be solved.
The orchestrator then can set the appropriate scalar inputs, and inquire for the scalar outputs, in the right order.
Please refer to \cite{Arnold2014} and \cite[Section 3.2]{Gomes2016a} for details on how to represent the graph, compute the topological sort, and identify the algebraic loops.

\subsubsection{Input/Output Couplings}

We assumed that the outputs and inputs of the simulation units are coupled by simple assignments.
In general, this might not be the case, as is shown in \cite{Schweizer2015}, and orchestration algorithms exists that deal with such advanced couplings. 
For example, see \cite[Section 4.3.1]{Gomes2017} and references thereof.

\subsubsection{Adaptive Communication Step}

We assumed that the communication step size has to remain fixed over the co-simulation.
In practice, it is often better that the orchestrator varies the communication step size, and/or asks the simulation units to adjust their own numerical method and approximation schemes, in order to respond to external requirements or react to the past dynamics of the co-simulation.

\subsection{Summary and Further Reading}

This section showed how the specialization of numerical methods naturally leads to the need for co-simulation.
Then it defined the intervenients in the co-simulation process and introduced the different kinds of orchestration algorithms.
Finally, it discusses some of the advanced techniques.

We refer the reader to \cite{Hafner2017,Schweiger2018a} for other motivations of co-simulation.
For in-depth discussions about the different kinds of orchestration algorithms and advanced techniques, please see \cite[Section 4]{Gomes2018} and references thereof.

\section{Conclusion}

The co-simulation technique introduced in the previous section allows us to apply the best numerical method to each part of a given IVP.
This is not the only benefit though.

For example, each numerical method can use a different step size.
This is an advantage because different models may evolve with derivatives that are orders of magnitude apart, as is 
the case in \cref{ex:car_driver} where the accelerations measured in the \unit{car} model are four orders of magnitude lower than the accelerations measured in the \unit{psg} unit.
Therefore, as shown in the co-simulation computed in \cref{fig:cosim_simple}, the \unit{car} unit can afford to take one simulation step per co-simulation step, while the \unit{psg} unit takes 100 steps, without drastically affecting the overall accuracy.

\claudio{Help wanted: Highlight the benefits in terms of model evaluations of the cosim in \cref{fig:cosim_simple}. And measure accuracy! So that we can see how much accuracy was lost/gained.}

Another benefit is that simulation units do not have to disclose the equations being solved internally.
Instead, it is common to only disclose the outputs and inputs, capabilities such as the ability to rollback, and the derivatives of outputs with respect to time and inputs.
The black box nature of the units makes it easier to standardize their interface, which in turn enables the coupling of mature modeling and simulation tools.
Wide industrial adoption is one of the main drivers behind research into co-simulation \cite{Schweiger2018}.

This tutorial aims at introducing the main concepts in co-simulation, and providing researchers and practitioners with further reading in each of the topics.
The concepts introduced here represent the fundamental concepts in co-simulation, general to any co-simulation framework.

\section*{Acknowledgments}

This work was executed under the framework of the COST Action IC1404 -- Multi-Paradigm Modelling for Cyber-Physical Systems (MPM4CPS), and partially supported by: Flanders Make vzw, the strategic research centre for the manufacturing industry; the INTO-CPS project funded by the European Commission's Horizon 2020 programme under grant agreement number 664047; and PhD fellowship grants from the Agency for Innovation by Science and Technology in Flanders (IWT, dossier 151067).

We thank Mehrdad Moradi for his comments and suggestions regarding this document.

\appendix

\section{Numerical Stability}
\label{sec:numerical_stability}

In this section, we introduce the concept of stability of a system of ordinary differential equations, and derive the conditions under which the numerical methods introduced in \cref{sec:sim_units} preserve this property.

We say that the system of differential equations in \cref{eq:ivp} is \emph{asymptotically stable} when all its solutions tend to zero as time passes, regardless of the initial value.
Formally, $\lim_{t \to \infty} \norm{\vect{x}(t)} = 0$ for all $x(t)$ satisfying \cref{eq:ivp}.

The following example illustrates why asymptotical stability is an important property.

\begin{example}
  Consider the solution $v(t)$ of the cruise controlled car IVP, introduced in \cref{ex:car}.
  
  After some time, the velocity of the car will be constant. Let $v_t$ denote this velocity.
  It can be computed by noting that the acceleration of the car will be zero at that speed.
  Hence, setting the right hand side of \cref{eq:car_ivp} to 0, and rearranging gives 
  $v_t = (k v_d) / (k + c_f)$.
  
  Determining the stability of \cref{eq:car_ivp} allows us to prove that the velocity of car actually tends to $v_t$, which is an important property of the cruise controller.
  Let $a = -(1/m)(k+c_f)$ and $b = (1/m)(k v_d)$, so that \cref{eq:car_ivp} can be written as $\dert{v} = av + b$, and $v_t = -b/a$.
  Then introduce a new variable $\bar{v}=v-v_t$ representing the difference between the car velocity and the terminal velocity.
  With the new variable, \cref{eq:car_ivp} can be written as $\dert{\bar{v}} = a\bar{v}$.
  Since $a < 0$, any solution $\bar{v}(t) \to 0$ as $t \to \infty$, independently of $\bar{v}(0)$, thus proving that the cruise controller is asymptotically stable\footnote{Notice that $v_t < v_d$ for $k>0$ and $d>0$. This makes the cruise controller incorrect, but keeps the example simple.}.
\end{example}

As the previous example shows, a scalar ODE in the form of 
\begin{equation}\label{eq:scalar_linear_ode}
\dert{x} = ax
\end{equation}
is asymptotically stable if $a<0$.
The analogous condition for vector ODEs of the form of
\begin{equation}\label{eq:linear_ode}
\dert{\vect{x}} = A\vect{x}, \text{ with $A$ being a constant matrix,}
\end{equation}
 is that the real part of all eigenvalues of $A$ is strictly negative \cite[Section I.12]{Wanner1991}.
Formally,
\begin{equation}
\forall \lambda \in \eig{A},\ \real{\lambda} < 0.
\end{equation}
Both these conditions can be checked automatically.

To see why a numerical method may fail to preserve the asymptotic stability of a system of differential equations, consider a scalar ODE in the form of \cref{eq:scalar_linear_ode}, and apply the explicit Euler method (\cref{eq:euler_method_matrix}) to get 
$x(t+h) \approx x(t) + a h x(t) = (1 + a h) x(t) = (1 + a h)^{n} x(0)$, where $n=t/h$.
For any $x(0)$, the term $(1 + a h)^{n} x(0) \to 0$ as $n \to \infty$ if $\abs{1 + a h} < 1$.
Note that the larger $\abs{a}$ is, the smaller the step size has to be, in order for the method to be numerically stable.

The vector version of the above derivation is analogous.
Consider a vector ODE in the form of \cref{eq:linear_ode}, and apply the explicit Euler method (\cref{eq:euler_method}) to get
$\vect{x}(t+h) \approx (I + A h)^{n} \vect{x}(0)$, where $n=t/h$.
For any $\vect{x}(0)$, the term $(I + A h)^{n} \vect{x}(0) \to 0$ as $n \to \infty$ if $\rho(I + A h) < 1$ \cite{Strang1993}, where $\rho(\cdot)$ denotes the maximum absolute eigenvalue of $\cdot$, also called the spectral radius of $\cdot$.
The analogous condition for the implicit Euler is $\rho((I + A h)^{-1}) < 1$, where $M^{-1}$ is matrix inverse of $M$.

The procedure to decide the \emph{numerical stability} is summarized as follows.
Apply the equation representing the numerical approximation to a differential equation of the form of \cref{eq:linear_ode}, and obtain an equation with the form
\begin{equation}\label{eq:dts}
\vect{x}(t+h) \approx \tilde{A} \vect{x}(t)
\end{equation}
where $\tilde{A}$ is a constant matrix.
Then check whether $\rho(\tilde{A}) < 1$.

When the explicit Euler is numerically unstable, a solution is to decrease the step size $h$, as it decreases the quantity $\rho(I + A h)$.
However, as the next example shows, the step size required to obtain a stable solution can be prohibitively small.
When this is the case, we recommend the use of a different numerical method, with better stability properties, such as the implicit Euler method.

\begin{example}\label{ex:msd_simple}
  Consider the mass-spring-damper, introduced in \cref{ex:msd}, with $m=1$, $f_e(t) = 0$ for any $t$, $0 < c_f < 1$, and $c^2 > 1$.
  It can be written in the form of \cref{eq:linear_ode}, with 
  \begin{equation*}
  A = \vectorTwo{0 & 1}{-c^2 & -c_f} \text{, where } 0<c_f<1, c>1.
  \end{equation*}
  The numerical stability of the explicit Euler method with the above equation is determined by $0.5 \abs{-c_fh + h \sqrt{c_f^2 - 4c^2} + 2} < 1$.
  For $0<h<1$, this inequality can be simplified to highlight the real and imaginary parts of the left hand side,
  $$0.5 \abs{-c_f h  + 2 + h \sqrt{c_f+2c}\sqrt{2c-c_f}\sqrt{-1}} < 1.$$
  Computing the absolute and simplifying gives
  $\abs{c^2 h^2 - c_f h + 2} < 1$.
  As the parameter $c_f \to 0$, the maximum safe step size $h \to 0$ as well, which means that in the limit where $c_f=0$, the explicit Euler method will never preserve the stability property.
  The same can be observed as the parameter $c \to \infty$.
\end{example}

\claudio{Help wanted: apply stability analysis for the implicit euler method, just like it is done for the explicit euler method above.}

\subsection{Further reading}

Other definitions of stability are given in \cite[Section 2.3]{Stuart1998}.
The derivation of the conditions for stability of vector ODE's is taken from \cite[Section I.12]{Wanner1991} and \cite{MIT2009a}.
For the stability of adaptive numerical methods (e.g., ones that change the step size over time), we refer to \cite{Gomes2017d,Gomes2018c,Gomes2018d}.

\section{Approximation Accuracy}
\label{sec:convergence}

In the previous section, we looked at whether a qualitative property of the original IVP could be preserved under a numerical simulation.
In this sub-section, we start by introducing a technique to compute the correct solution to a restricted class of differential equations, so that we can later show how to experimentally compute the approximation error of a numerical method, as a function of the step size used.

\subsection{Analytical Solution}
\label{sec:analytical_solution}

When the IVP in \cref{eq:ivp} has the form of \cref{eq:linear_ode}, the \emph{analytical solution} is given by
\begin{equation}\label{eq:solution_linear_ode}
\vect{x}(t)=e^{At}\vect{x}_0 \text{, with } e^{At} = I + At + \frac{A^2 t^2}{2!} + \frac{A^3 t^3}{3!} + \ldots,
\end{equation}
where $e^{At}$ is the matrix exponential of $At$ \cite[Section 2.2]{DerekRowell2002}.
This can be verified by taking the derivative of $e^{At}\vect{x}_0$, and obtaining $A e^{At}\vect{x}_0 = A \vect{x}$ (the right hand side of the IVP in \cref{eq:ivp}).

Most software libraries include algorithms to approximate the matrix exponential in \cref{eq:solution_linear_ode}.
Furthermore, the computation of $\vect{x}(t)$ can be done incrementally, mimicking a numerical method, by noting that 
$\vect{x}(t+h) = e^{A(t+h)}\vect{x}_0 = e^{Ah}e^{At}\vect{x}_0 = e^{Ah}\vect{x}(t)$.

The formulation in \cref{eq:solution_linear_ode} is generic enough to allow the computation of the solution to equations of the form 
\begin{equation}\label{eq:linear_forced_ode}
\dert{\vect{x}} = A \vect{x} + \vect{b}
\end{equation}
where $\vect{b}$ is a constant vector:
\cref{eq:linear_forced_ode} can be transformed to the form in \cref{eq:linear_ode} by introducing a new state vector $\vect{\hat{x}} = \vectorOne{ x_1 & \ldots & x_n & u}^T$ and solving the IVP
\begin{equation*}
\dert{\vect{\hat{x}}} = 
\left[ \begin{array}{c|c}
A & \vect{b} \\
\hline 
\vect{0}_{1 \times n} & \vect{0}_{1 \times 1}
\end{array}
\right]
\vect{\hat{x}} \text{, with } \vect{\hat{x}}(0)=
\left[ \begin{array}{c}
\vect{x_0} \\
\hline 
1
\end{array}
\right],
\end{equation*}
where $\vect{0}_{p \times q}$ denotes the null matrix with dimensions $p \times q$.

\subsection{Experimental Approximation Error}

Given an approximation $\tilde{\vect{x}}(t)$ of the solution $\vect{x}(t)$ to the IVP introduced in \cref{eq:ivp}, we define the approximation error of an approximation computed with step size $h$ as $\vect{e}_h (t) = \vect{x}(t) - \tilde{\vect{x}}(t)$, and the maximum error up to a finite $T>0$ as $e_{\mathop{max} T}(h)=\max_{i \in \set{0, \ldots, T/h}} \norm{\vect{e}_h(ih)}$.

We can experimentally plot the error $e_{\mathop{max} T}(h)$ of a numerical method applied to an IVP as a function of the step size $h$ as follows. Pick a finite simulation time $T>0$; and compute the maximum error $e_{\mathop{max} T}(h)$ of the analytical and numerical solutions up to $T$ for different step sizes.

The resulting $e_{\mathop{max} T}(h)$ can be used to get the \emph{order} of the numerical method.
Roughly, for sufficiently small $h$, if $e_{\mathop{max} T}(h) < c \abs{g(h)}$ for a given function $g(h)$ and positive constant $c$, then we say 
that $e_{\mathop{max} T}(h)$ is in the order of $g(h)$, or in other words, $e_{\mathop{max} T}(h) = \bigO{g(h)}$.
The constant $c$ that approximates the error depends on the IVP being solved, but the order is a property of the numerical method \cite[Section I.7 and II.1]{Wanner1991}.

\begin{figure}[tbh]
\begin{center}
  \includegraphics[width=0.7\textwidth]{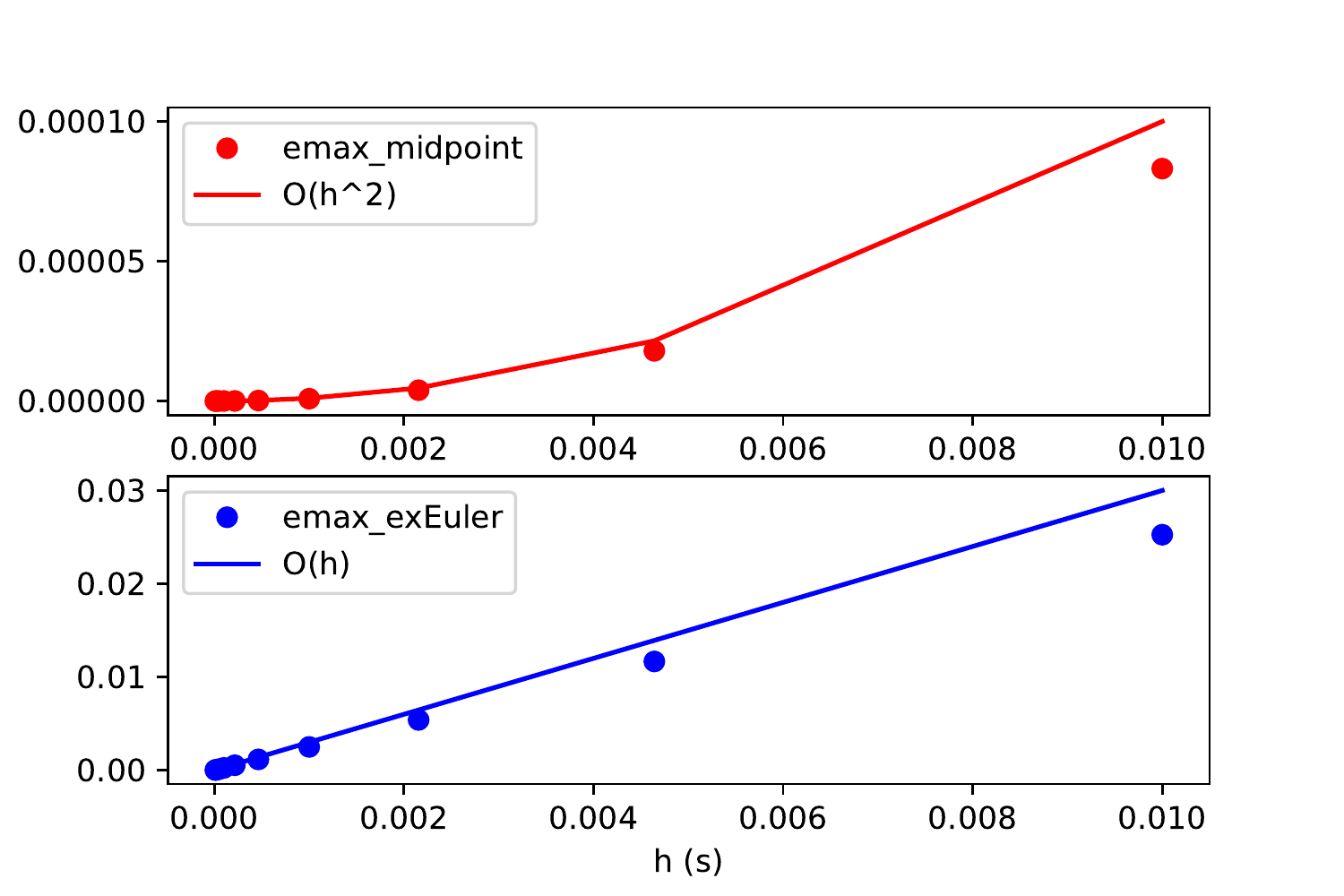}
  \caption{Approximation order of the explicit Euler and the Midpoint methods.}
  \label{fig:convergence_euler_vs_midpoint}
\end{center}
\end{figure}

\begin{example}

The Midpoint method is given by the iteration:
\begin{equation*}
\vect{x}(t+h) \approx \vect{x}(t) + F(\vect{x}(t) + F(\vect{x}(t),\vect{u}(t)) 0.5 h,\vect{u}(t + 0.5 h)) h \text{, with } \vect{x}(0)=\vect{x_0}.
\end{equation*}
\Cref{fig:convergence_euler_vs_midpoint} compares the approximation error of the explicit Euler method (\cref{eq:euler_method_matrix}) with the approximation error of the Midpoint method, when applied to the IVP introduced in \cref{ex:msd_simple}.
The Midpoint method is $\bigO{h^2}$, while the explicit Euler is $\bigO{h}$.
Also note that, for the same maximum error, the step size required by the Midpoint method is larger than the required for the explicit Euler.
\end{example}

\subsection{Further Reading}

We refer the reader to \cite{Cellier2006} and \cite{Wanner1991} for theorical discussions of error in numerical methods.

\section{Co-simulation Convergence and Stability}

The methods introduced in \cref{sec:numerical_stability,sec:convergence} can be applied to co-simulations as well.

\subsection{Stability}

To study the numerical stability of a co-simulation, one: 
\begin{compactenum}
\item starts with a linear system in the form of \cref{eq:linear_ode},
\item describes each simulation unit with an equation as in \cref{tab:types_sim_units},
\item couples the multiple simulation units, and
\item ends with an iteration of the form
$\vect{\tilde{x}}(t+H) \approx \tilde{A} \vect{\tilde{x}}(t)$,
representing the co-simulation method.
\end{compactenum}
The state vector $\vect{\tilde{x}}$ is the concatenation of the state vectors of each simulation unit, and the matrix $\tilde{A}$ encodes every action of every simulation unit to complete each step of the co-simulation.
The condition $\rho(\tilde{A}) < 1$ can then be checked.

We now exemplify this procedure for the co-simulation of two simulators, connected in a feedback loop, without algebraic loops.
This procedure can be generalized to any number of simulators, as long as the underlying coupled system can be written as in \cref{eq:linear_ode} (for conditions that ensure this, see \cite[Section 2]{Arnold2014}).
We consider two orchestration methods: a Jacobi and an Iterative Jacobi scheme.

\subsubsection{Jacobi Orchestration}
\label{sec:jacobi_derivation}

Time is discretized into a countable set $T = \set{t_0, t_1, t_2, \ldots} \subset \setreal$, where $t_{i+1} = t_i + H_i$ is the time at step $i$ and $H_i$ is the communication step size at step $i$, with $i=0,1,\ldots$

In the interval $t \in \brackets{t_i, t_{i+1}}$, each simulator $S_j$ approximates the solution to a linear ODE,
\begin{aligneq}\label{eq:linear_ode_inputs}
\dert{\vect{x}}_j &= A_j  \vect{x}_j + B_j  \vect{u}_j \\
\vect{y}_j    &= C_j  \vect{x}_j + D_j  \vect{u}_j
\end{aligneq}
where $A_j, B_j, C_j, D_j$ are matrices, the initial state $\vect{x}_j(t_i)$ is computed in the most recent co-simulation step, and $j=1,2$.

Since the simulators only exchange outputs at times $t_i, t_{i+1} \in T$, the input $\vect{u}_j$ has to be extrapolated in the interval $[t_i, t_{i+1})$.
In the simplest co-simulation strategy\footnote{The derivation presented can be applied to more sophisticated input extrapolation techniques, see \cite[Equation~(9)]{Busch2016}.}, this extrapolation is often implemented as a zero-order hold:
$\tilde{\vect{u}}_j(t) = \vect{u}_j(t_i)$, for $t \in [t_i, t_{i+1})$.
Then, \cref{eq:linear_ode_inputs}
can be re-written to represent the unforced system being integrated by each simulator:
\begin{aligneq}\label{eq:linear_ode_unforced}
\vectorTwo{\dert{\vect{x}}_j}{\dert{\tilde{\vect{u}}}_j} &= \vectorTwo{A_j & B_j}{\mathbf{0} & \mathbf{0}}  \vectorTwo{\vect{x}_j}{\tilde{\vect{u}}_j}
\end{aligneq}

We can represent the multiple internal integration steps of \cref{eq:linear_ode_unforced}, performed by the simulator $S_j$ in the interval $t \in \brackets{t_i, t_{i+1}}$, as
\begin{aligneq}\label{eq:solver_internal_steps}
\vectorTwo{\tilde{\vect{x}}_j(t_{i+1})}{\tilde{\vect{u}}_j(t_{i+1})} &= \tilde{A}^{k_j}_j  \vectorTwo{\tilde{\vect{x}}_j(t_i)}{\tilde{\vect{u}}_j}
\end{aligneq}
where, e.g., $\tilde{A}_j = \mathbf{I} + h_j \vectorTwo{A_j & B_j}{\mathbf{0} & \mathbf{0}}$ for the Forward Euler method,
$k_j=(t_{i+1} - t_i)/h_j$ is the number of internal steps, 
and $0 < h_j \leq H_i$ is the internal fixed step size that divides $H_i$.

We assumed that the two simulators are coupled in a feedback loop:
\begin{equation}\label{eq:couplings}
\vect{u}_1 = \vect{y}_2 \text{ and } \vect{u}_2 = \vect{y}_1,
\end{equation}
and that there are no algebraic loops, so either $D_1$ or $D_2$ is the zero matrix. 
Let $D_2 = \mathbf{0}$. 

With the Jacobi orchestration algorithm (recall \cref{alg:ct_cosim_jacobi}), at the beginning of the co-simulation step $i$, $\vect{u}_1(t_i) = \vect{y}_2(t_i)$ and $\vect{u}_2(t_i) = \vect{y}_1(t_i)$.
This, together with \cref{eq:linear_ode_inputs}, gives,
\begin{aligneq}\label{eq:io_couplings}
\vect{u}_1(t_i) &= C_2  \tilde{\vect{x}}_2(t_i) \\
\vect{u}_2(t_i) &= C_1  \tilde{\vect{x}}_1(t_i) + D_1 C_2  \tilde{\vect{x}}_2(t_i).\\
\end{aligneq}
\Cref{eq:linear_ode_unforced,eq:solver_internal_steps,eq:io_couplings} can be used to represent each co-simulation step in the form of \cref{eq:dts}:
{\smaller
\begin{aligneq}
\notag
\vectorTwo{\tilde{\vect{x}}_1(t_{i+1})}{\tilde{\vect{x}}_2(t_{i+1})} &=
  \underbrace{
  \vectorTwo{\mathbf{I} & \mathbf{0} & \mathbf{0} & \mathbf{0}}{\mathbf{0} & \mathbf{0} & \mathbf{I} & \mathbf{0}}
  \vectorTwo{\tilde{A}^{k_1}_1 & \mathbf{0}}{\mathbf{0} & \tilde{A}^{k_2}_2}
  \vectorFour{\mathbf{I} & \mathbf{0}}{\mathbf{0} & C_2}{\mathbf{0} & \mathbf{I}}{C_1 & D_1  C_2}
  }_{\tilde{A}}
  \vectorTwo{\tilde{\vect{x}}_1(t_{i})}{\tilde{\vect{x}}_2(t_{i})}
\end{aligneq}
}

\subsubsection{Iterative Jacobi Orchestration}

Here we assume that each co-simulation unit is represented in the form:
\begin{equation}\label{eq:cosim_unit}
\vectorTwo{\tilde{\vect{x}}_j(t_i + H)}{\tilde{\vect{u}}_j(t_i + H)} = \vectorTwo{M_{1,\vect{x}_j} & M_{1,\vect{u}_j}}{M_{2,\vect{x}_j} & M_{2,\vect{u}_j}}  \vectorTwo{\tilde{\vect{x}}_j(t_i)}{\vect{u}_j(t_i)}
\end{equation}
The derivation of which is explained in \cref{sec:jacobi_derivation}.

In the iterative Jacobi method (recall \cref{fig:cosim_overview}), at the beginning of the co-simulation step $i+1$, there is a successive substitution fixed point iteration.
This can be modelled by
\begin{aligneq}
\vect{u}_1(t_{i+1}) &= C_2  \tilde{\vect{x}}_2(t_{i+1}) \\
\vect{u}_2(t_{i+1}) &= C_1  \tilde{\vect{x}}_1(t_{i+1}) + D_1 \vect{u}_1(t_{i+1}) .
\end{aligneq}

As done in \cref{sec:jacobi_derivation}, the above equation can be expanded and simplified to:
\begin{aligneq}
\tilde{\vect{x}}_1(t_{i+1}) &= M_{1,x_1} \tilde{\vect{x}}_1(t_i)  +  M_{1,\vect{u}_1} C_2  \tilde{\vect{x}}_2(t_{i+1}) \\
\vect{u}_1(t_{i+1}) &= M_{2,x_1}  \tilde{\vect{x}}_1(t_i)  +  M_{2,\vect{u}_1} C_2  \tilde{\vect{x}}_2(t_{i+1})  \\
\tilde{\vect{x}}_2(t_{i+1}) &=   M_{1,x_2} \tilde{\vect{x}}_2(t_i)  +  M_{1,\vect{u}_2} C_1  \tilde{\vect{x}}_1(t_{i+1}) +  M_{1,\vect{u}_2} D_1 \vect{u}_1(t_{i+1})   \\
\vect{u}_2(t_{i+1})  &= M_{2,x_2}  \tilde{\vect{x}}_2(t_i)  +  M_{2,\vect{u}_2} C_1  \tilde{\vect{x}}_1(t_{i+1}) + M_{2,\vect{u}_2} D_1 \vect{u}_1(t_{i+1}) 
\end{aligneq}
which can be put in matrix form:
\begin{aligneq}
\vectorFour{\tilde{\vect{x}}_1(t_{i+1})}%
      {\vect{u}_1(t_{i+1})}%
      {\tilde{\vect{x}}_2(t_{i+1})}%
      {\vect{u}_2(t_{i+1})} &=  \vectorFour{M_{1,x_1} & 0 & 0 & 0}%
                             {M_{2,x_1} & 0 & 0 & 0}%
                             {0 &  0 & M_{1,x_2} & 0}%
                             {0 & 0 & M_{2,x_2} & 0}
                    \vectorFour{\tilde{\vect{x}}_1(t_i)}%
                           {\vect{u}_1(t_i)}%
                           {\tilde{\vect{x}}_2(t_i)}%
                           {\vect{u}_2(t_i)} + \\ 
                      & \vectorFour{0 & 0 & M_{1,\vect{u}_1} C_2 & 0}%
                             {0 & 0 & M_{2,\vect{u}_1} C_2 & 0}%
                             {M_{1,\vect{u}_2} C_1 & M_{1,\vect{u}_2} D_1 & 0 & 0}%
                             {M_{2,\vect{u}_2} C_1 & M_{2,\vect{u}_2} D_1 & 0 & 0}
                    \vectorFour{\tilde{\vect{x}}_1(t_{i+1})}%
                           {\vect{u}_1(t_{i+1})}%
                           {\tilde{\vect{x}}_2(t_{i+1})}%
                           {\vect{u}_2(t_{i+1})}
\end{aligneq}

Renaming the above equation to $\bar{\vect{x}}_{i+1} =  \bar{M}_i \bar{\vect{x}}_i +  \bar{M}_{i+1} \bar{\vect{x}}_{i+1}$, we get an equation in the form of \cref{eq:dts}:
\begin{aligneq}\label{eq:into_coupling}
\bar{\vect{x}}_{i+1} &=  (I -  \bar{M}_{i+1}) ^{-1} \bar{M}_i \bar{\vect{x}}_i 
\end{aligneq}

\subsection{Convergence}

Regarding the accuracy of the co-simulation, the analysis is more difficult, but not fundamentally different than the one introduced here.
The added difficulty arises from the fact that, besides the numerical methods employed by each unit and their internal step size,
the communication step size and the input approximation functions, also have to be taken into account.
The combination of parameters makes it hard to judge the accuracy of the co-simulation.

\subsection{Further Reading}

The work in \cite[Section 4]{Gomes2018} provides an overview of references that focus on the stability of co-simulation methods.
The theoretical foundations for the stability of adaptive orchestration algorithms are discussed in \cite{Gomes2017d,Gomes2018c,Gomes2018d}.
Regarding the convergence of co-simulation methods, we refer to \cite{Kubler2000,Arnold2014}.

\bibliographystyle{plain}
\bibliography{bibliography_claudio}

\end{document}